\documentclass[11pt,letterpaper]{article}
\usepackage[margin=1in]{geometry}
\usepackage[T1]{fontenc}
\usepackage{lmodern,microtype,amsmath,amssymb,amsthm,enumitem}
\usepackage{xcolor,tikz,float}
\usetikzlibrary{arrows.meta,calc,positioning}
\definecolor{ecA}{HTML}{D55E00}
\definecolor{ecB}{HTML}{0072B2}
\definecolor{ecC}{HTML}{009E73}
\definecolor{ecD}{HTML}{8C6BB1}
\definecolor{ecE}{HTML}{C99A00}
\floatstyle{ruled}
\newfloat{algorithm}{htbp}{loa}
\floatname{algorithm}{Algorithm}
\newlist{steps}{enumerate}{1}
\setlist[steps]{label=\textcolor{black!55}{\scriptsize\arabic*},leftmargin=2em,
  labelsep=.65em,itemsep=1.5pt,topsep=4pt,parsep=0pt}
\newcommand{\st}[2][0]{\item\begingroup\advance\leftskip by #1em\relax #2\par\endgroup}
\newcommand{\io}[2]{\small\noindent\textbf{Input}\quad #1\par
  \noindent\textbf{Output}\quad #2\par}
\usepackage[numbers,sort&compress]{natbib}

\usepackage[hidelinks]{hyperref}
\usepackage{bookmark}
\newtheorem{theorem}{Theorem}
\newtheorem{lemma}{Lemma}[section]
\newtheorem{proposition}[lemma]{Proposition}
\newcommand{\softO}{\widetilde O}
\newcommand{\E}{\mathbb E}
\newcommand{\F}{\mathbb F}
\newcommand{\ceil}[1]{\lceil #1\rceil}
\newcommand{\floor}[1]{\lfloor #1\rfloor}
\setlist{itemsep=2pt,topsep=4pt}
\hypersetup{pdftitle={Parallel Edge Coloring in Square-Root Depth},
  pdfauthor={Jeff Giliberti, MohammadTaghi Hajiaghayi, Changki Yun}}
\title{Parallel Edge Coloring in Square-Root Depth}
\author{Jeff Giliberti \qquad MohammadTaghi Hajiaghayi \qquad Changki Yun\\[3pt]
  \normalsize University of Maryland, College Park\\
  \small\texttt{\{jeffgili,hajiagha,ckyun\}@umd.edu}}
\date{}
\begin{document}
\maketitle

\begin{abstract}
We give a deterministic parallel algorithm that finds a $(\Delta+1)$-edge coloring of an $n$-vertex graph with maximum degree $\Delta$ in $\tilde O(\min\{\Delta,\sqrt n\})$ depth using $O(n^3\Delta)$ processors. Achieving sublinear depth across all maximum degrees had remained open since the seminal work of Karloff and Shmoys~[J.~Algorithms~'87]. In terms of $\Delta$, this improves upon the recent work of Fineman and Mohammadi~[arXiv~'26], who obtained $\tilde O(\Delta^2)$ randomized and $\tilde O(\Delta^3)$ deterministic depth.

Our algorithm recursively removes $\tilde O(n/\Delta)$ vertices to lower the maximum degree by two and splits the remaining edges along Euler tours into two subgraphs colored in parallel on disjoint palettes. To reinsert the removed stars, we rotate fans before finding finishing paths, tracing alternating paths from both ends to choose rotations deterministically. Restoring stars either one by one in $\tilde O(n/\Delta)$ depth or in independent groups in $\tilde O(\Delta)$ depth gives the final bound.
\end{abstract}

\section{Introduction}
A proper edge coloring of a graph assigns colors to its edges such that incident edges receive different colors. For a graph with maximum degree $\Delta$, every edge coloring requires at least $\Delta$ colors. Vizing's theorem~\cite{Vizing64} guarantees that $\Delta+1$ always suffice. The algorithmic question is how efficiently such a coloring can be found.

In the sequential setting, Vizing's proof gives an $O(mn)$-time algorithm on a graph with $n$ vertices
and $m$ edges. Arjomandi~\cite{Arj82} and Gabow et al.~\cite{GNKLT85}
 reduced the sequential time to $\softO(m\sqrt n)$\footnote{
$\softO$ suppresses polylogarithmic factors.}. A sequence of randomized improvements culminated in
a near-linear time algorithm by Assadi, Behnezhad,
Bhattacharya, Costa, Solomon, and Zhang~\cite{ABBCSZ26R}.
The same authors subsequently obtained a deterministic algorithm
with almost-linear running time $m^{1+o(1)}$~\cite{ABBCSZ26D}. 
While these results bring the sequential running time close to linear, they do
not reduce the parallel depth, i.e., the longest chain of dependent operations.

Parallel edge coloring has been studied since the early 1980s.
Lev, Pippenger, and Valiant~\cite{LPV81} gave a polylogarithmic-depth
algorithm that colors bipartite graphs with exactly $\Delta$ colors.
Karloff and Shmoys~\cite{KS87} then gave a deterministic parallel
$(\Delta+1)$-edge-coloring algorithm with $\softO(\Delta^7)$ depth.
Subsequent work steadily reduced the polynomial dependence on $\Delta$.
Liang, Shen, and Hu~\cite{LSH96} achieved depth $\softO(\Delta^{4.5})$ under
the corrected analysis of Elkin and Khuzman~\cite{EK26}, who themselves
obtained deterministic depth $\softO(\Delta^4)$.
Most recently, Fineman and Mohammadi~\cite{FM26} lowered the depth to
$\softO(\Delta^3)$ deterministically and $\softO(\Delta^2)$ with randomization.
For unrestricted degree, finding a polylogarithmic-depth algorithm
with polynomially many processors is a longstanding open
problem, which appears in the book of Greenlaw, Hoover, and
Ruzzo~\cite[Problem B.9.3]{GHR95}.

These $\Delta$-dependent bounds widen the range of degrees admitting fast
parallel algorithms, yet remain at least linear in $n$ once
$\Delta = \Omega(\sqrt{n})$. Despite nearly four decades of work on parallelizing Vizing's theorem,
achieving sublinear depth $o(n)$ for unrestricted $\Delta$ has remained open.

\subsection{Contribution}
Our main result is the first sublinear-depth parallel algorithm for $(\Delta+1)$-edge coloring.
\begin{theorem}\label{thm:main}
    There is a deterministic uniform CREW PRAM algorithm that finds a $(\Delta+1)$-edge
    coloring of an $n$-vertex graph of maximum degree $\Delta$ in
    $\softO(\min\{\Delta,\sqrt n\})$ depth using $O(n^3\Delta)$ processors.
\end{theorem}
Theorem~\ref{thm:main} reduces the degree dependence from $\softO(\Delta^3)$ deterministically and $\softO(\Delta^2)$ with randomization in~\cite{FM26} to deterministic $\softO(\Delta)$.
At the same time, it caps worst-case depth at $\softO(\sqrt{n})$.

It is worth noting that sublinear depth does not imply NC membership. For example, Square Circuit Value is P-complete but admits $\softO(\sqrt{s})$ depth for circuits of size $s$~\cite[Theorem 9.2.7]{GHR95}. Whether $(\Delta+1)$-edge coloring admits polylogarithmic depth for unrestricted degree remains open.

\section{Technical overview}\label{sec:overview}
Our algorithm combines two main building blocks: a recursive split and a
star-restoration routine. At each recursive level, we remove a small
balanced set of vertices to lower the maximum degree by two, divide the
remaining graph along Euler tours into two parallel subproblems on
disjoint palettes, and then reinsert the removed vertices. We begin with
a single uncolored edge and show how local fan rotations move it to a
better position while keeping two ways to avoid an obstructing
alternating path (Section~\ref{sec:ov-edge}). Next, we scale this
operation to color an entire star around one vertex in polylogarithmic
depth, and then extend it to an independent set of stars whose
neighborhoods have small overlap (Section~\ref{sec:ov-stars}). To run
many fans together deterministically, we rotate the chosen fans first
and resolve their alternating paths afterward, so that a constant
fraction of the rotated fans finish together
(Section~\ref{sec:ov-rotations}). We then combine both restoration modes
with the recursive split, reinserting removed vertices one by one when
$\Delta$ is large and in independent groups when $\Delta$ is smaller, to
obtain the $\softO(\min\{\Delta,\sqrt{n}\})$ depth bound
(Section~\ref{sec:ov-sqrt}). We conclude this section by comparing our approach with prior
work in more detail (Section~\ref{sec:ov-compare}).

\subsection{Moving an uncolored edge}\label{sec:ov-edge}
To color an uncolored edge, we rotate a local fan until a two-color swap
can color it, keeping two stopping points so the step survives other
rotations. Consider an uncolored edge $sx$, where we call $s$ the
center, its incident edges spokes, and their other endpoints leaves. If
$s$ and $x$ miss a common color, we color $sx$ immediately. Otherwise, a
color missing at $x$ must appear on a unique spoke at $s$. Moving that
color onto $sx$ leaves that spoke uncolored while keeping the same
colors at $s$. Repeating this step forms a fan, and any prefix of it
can be rotated in one parallel step once we know which spoke supplies
each color.

What is a useful place to stop? Suppose the center misses red and the
new uncolored spoke ends at a leaf missing blue. The red and blue edges
form paths and cycles; if the center and leaf lie in different
components, swapping red and blue on the leaf's component makes both
endpoints miss red, coloring the spoke. The only obstruction is a
red--blue path between them. Finding even a long path is fast in shared
memory; the difficulty is keeping an unobstructed leaf while other fans
rotate.

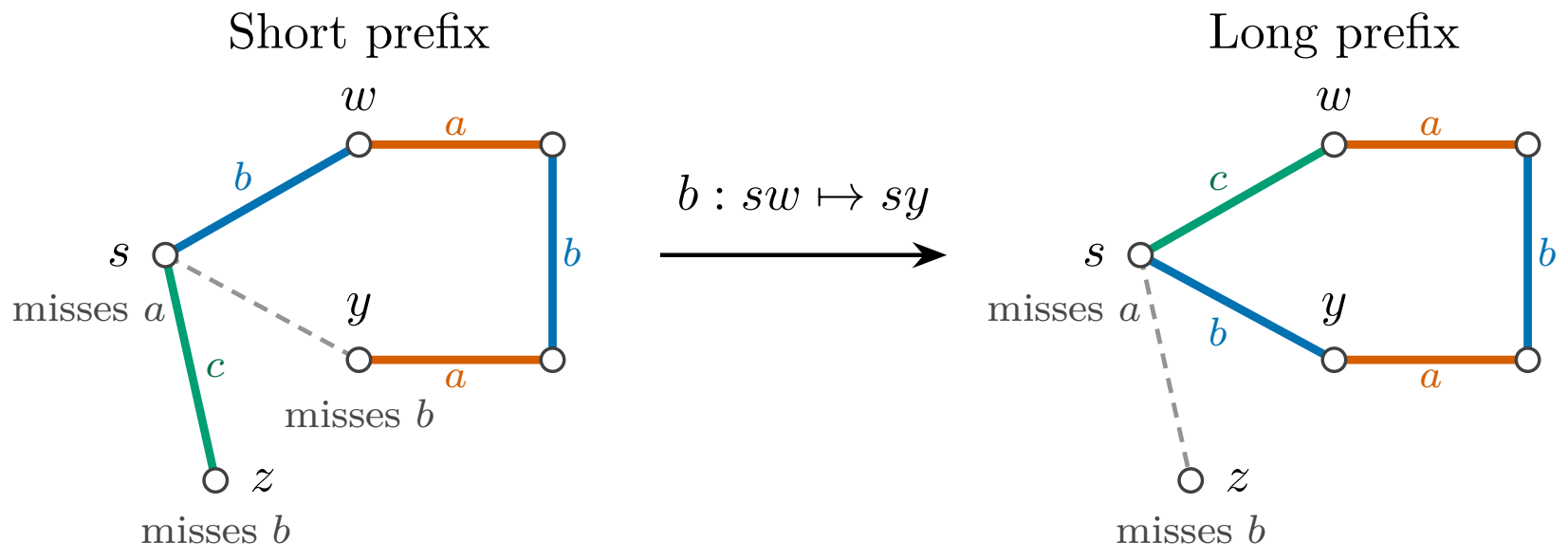
\begin{figure}[ht]
\centering
% Nested fan: moving the b-spoke from sw to sy preserves the vertex set
% of s's (a,b)-component and leaves the long-prefix blank sz outside it.
% The pointer cycle w,z has length two; ecC is the third color c.
\begin{tikzpicture}[
  x=1cm,y=1cm,font=\small,>=Stealth,
  v/.style={circle,draw=black!75,fill=white,inner sep=0pt,
            minimum size=4.6pt,line width=.65pt},
  e/.style={line width=1.65pt,line cap=round},
  blank/.style={draw=black!42,densely dashed,line width=.9pt},
  lab/.style={font=\scriptsize,inner sep=1.2pt},
  miss/.style={font=\scriptsize,text=black!72}
]
\foreach \panel/\shift/\heading in {0/0/Short prefix,1/6.8/Long prefix} {
  \begin{scope}[xshift=\shift cm]
    \node[font=\small] at (1.35,1.58) {\heading};
    \coordinate (s) at (0,0.05);
    \coordinate (w) at (1.35,0.82);
    \coordinate (p) at (2.70,0.82);
    \coordinate (q) at (2.70,-0.68);
    \coordinate (y) at (1.35,-0.68);
    \coordinate (z) at (0.35,-1.52);

    % Unchanged (a,b)-subpath from w to y.
    \draw[e,ecA] (w)--node[lab,above,text=ecA] {$a$} (p);
    \draw[e,ecB] (p)--node[lab,right,text=ecB] {$b$} (q);
    \draw[e,ecA] (q)--node[lab,below,text=ecA] {$a$} (y);

    \ifnum\panel=0
      \draw[e,ecB] (s)--node[lab,above left,text=ecB] {$b$} (w);
      \draw[e,ecC] (s)--node[lab,right=1pt,text=ecC!70!black] {$c$} (z);
      \draw[blank] (s)--(y);
      \node[miss,below=4pt] at (y) {misses $b$};
    \else
      \draw[e,ecB] (s)--node[lab,below left,text=ecB] {$b$} (y);
      \draw[e,ecC] (s)--node[lab,above left,text=ecC!70!black] {$c$} (w);
      \draw[blank] (s)--(z);
    \fi

    \foreach \pt in {s,w,p,q,y,z} {\node[v] at (\pt) {};}
    \node[left=3pt] at (s) {$s$};
    \node[miss,below left=4pt and -4pt] at (s) {misses $a$};
    \node[above=3pt] at (w) {$w$};
    \node[above=3pt] at (y) {$y$};
    \node[right=3pt] at (z) {$z$};
    \node[miss,below=3pt] at (z) {misses $b$};
  \end{scope}
}
\draw[->,line width=.85pt] (3.45,0.05)--node[above=3pt] {$b: sw \mapsto sy$} (5.45,0.05);
\end{tikzpicture}
\caption{A nested fan with pointer cycle $w,z$. The short rotation leaves
$sy$ blank and is obstructed at $y$. The long rotation moves $b$ from $sw$ to $sy$
and $c$ from $sz$ to $sw$, leaving $sz$ blank.
The $(a,b)$-component of $s$ keeps the same vertex set, so $z$ stays
outside it.}
\label{fig:prefix}
\end{figure}

Two leaves missing blue avoid this obstruction: since the center misses
red and both leaves miss blue, the center's red--blue component is a
path that can reach at most one of them (as its far endpoint). We obtain
such a pair in two ways. First, if the leaves come from two uncolored
spokes with disjoint fans, we can rotate both and at least one leaf
stays outside the path. Second, a single uncolored spoke gives two
leaves when its fan runs into a cycle: we can stop just before the cycle
or rotate farther around it. If the shorter rotation gives an
obstructing path to the first leaf, the longer one moves the center's
blue spoke from the start of that path to its far end (giving the old
spoke a third color). The center's red--blue component keeps the same
vertices, so the second leaf stays outside it (Figure~\ref{fig:prefix}).
Either way, at least one leaf still works after any fixed rotations at
other, vertex-disjoint fans.

\subsection{From one star to many}\label{sec:ov-stars}
We now scale this operation from one uncolored edge to an entire star
around a single center, and then to an independent set of stars whose
neighborhoods have small overlap.

\paragraph{One center.}
When several spokes at a center are uncolored, their fans can compete
for the same colors and leaves. We mark one missing color at each
neighbor and point every leaf to the unique spoke carrying that color,
if any. This forms a graph with out-degree at most one, made of trees
and cycles. Stopping two walks just before they meet produces two spokes
whose final leaves share a mark, while a walk entering a cycle leaves
one spoke with two stopping points. Processing the trees and then the
cycles in this way gives leaf-disjoint choices proportional to the
number of uncolored spokes. Because a
center with many uncolored spokes also misses many colors, choices with
distinct terminal marks can each reserve a different missing color at
the center. All the resulting color pairs are disjoint, so these choices
can be colored in parallel even though they share the center.

To keep many choices with distinct terminal marks, we first spread the
marks across the palette by recoloring. For any two colors, swapping
them on an entire alternating path changes missing colors only at the
path's two endpoints, so each path contributes at most two marks to any
count we want to balance. Pairing colors one bit at a time spreads the
marks across every center's neighborhood. Keeping one choice per
terminal mark then loses only a polylogarithmic factor, so repeating
restores an entire star in polylogarithmic depth, regardless of its
degree.

\paragraph{Several centers.}
We can restore many centers together when they form an independent set
and every other vertex has few neighbors in it, measuring progress over
all uncolored spokes at once. Here each center marks one missing color
of its own and pairs it with each choice's terminal mark (keeping one
choice per pair). Picking a matching of disjoint color pairs then leaves
at most one fan active per center. Active fans at different centers may
still share a leaf, especially when they ask for the same pair and
activate together. Spreading center marks among centers that share a
neighbor keeps these conflicts small, letting us select fans with
disjoint local rotations. Their finishing paths, however, can still pass
through one another's fans.

\subsection{Choosing rotations before finding paths}\label{sec:ov-rotations}
Given a batch of vertex-disjoint fans, we rotate them and find
finishing paths in the new coloring. For each fan with two possible lengths, a
random bit picks a usable terminal with probability at least one half for
every fixed choice of the other bits, so many terminals are usable on
average. Making this choice deterministically is harder: to replace
independent bits with a small explicit list, we want to test a terminal
using only a few bit queries, yet its alternating path can pass through
many fans and inspect their bits along the way.

The two directions along the path behave differently. A nested fan's
center misses the first color of its own pair, so it can only be an
endpoint for that pair. For a different traced pair it can be an interior
vertex, with a fresh bit choosing between two outgoing spokes.
At a leaf, only the spoke to its center can change: if that
lookup depends on a fresh bit, one value enters the center and the other
ends the walk. For any bit-dependent spoke at an interior fan center
on the realized path, either the forward or reverse walk crosses it \emph{toward}
its center, passing a fresh test of probability one half. Passing many such tests is unlikely.
Although the far endpoint depends on the rotations, the reverse walk has
only $n$ possible starting vertices and two possible first colors.
Accounting for both directions shows that, up to a constant loss in
success probability, testing a terminal can stop after $\log_2 n+O(1)$
bit queries. A polynomial-size list of assignments preserves this
probability up to a small constant; we test the list in parallel batches
and keep the assignment with the most usable terminals.

Once the rotations are fixed, we resolve the remaining conflicts
exactly. For each active color pair, we add its chosen uncolored edges
to the edges already carrying those two colors. Since the chosen edges
form a matching and each endpoint misses one color of the pair, the
graph splits into paths and cycles. Alternating the two colors fills
every chosen edge on a path or even cycle, while an odd cycle forces us
to leave one unfilled. Because each chosen edge joins two distinct
original components and its endpoints miss opposite colors, every odd
cycle contains an odd number of chosen edges and never just one---so at
least three. Dropping one per odd cycle fills at least two thirds
of the chosen edges (Lemma~\ref{lem:fixed-pair}).

\subsection{From star restoration to the square root}\label{sec:ov-sqrt}
We place our two star-restoration routines inside a recursive split
to obtain $\softO(\min\{\Delta,\sqrt{n}\})$ depth. At a
recursive level of maximum degree $D\le\Delta$, we first select
$\softO(n/D)$ vertices whose removal lowers the maximum degree by at
least two while leaving every vertex with only polylogarithmically many
selected neighbors. Sampling explains why these bounds fit together:
picking each vertex at a polylogarithmic rate over $D$ expects
$\softO(n/D)$ vertices overall, several neighbors at every high-degree
vertex, and at most polylogarithmically many at every vertex. We
implement this deterministically by repeatedly halving the vertex set
while balancing all neighborhood counts.

\begin{figure}[tbp]
\centering
% One recursive level of Algorithm 1 on W_5 (D=5):
% delete S={s} so D_0=2 <= D-2; Euler-split H-S into H_1 (D_1=2) and
% H_2 (D_2=1) colored recursively on disjoint palettes P_1, P_2; combine
% on H-S; then restore S in the full palette P.
% Requires TikZ arrows.meta and colors ecA--ecE.
\begin{tikzpicture}[
  x=1cm,y=1cm,font=\small,>=Stealth,
  v/.style={circle,draw=black!75,fill=white,inner sep=0pt,
            minimum size=4.4pt,line width=.65pt},
  vsmall/.style={circle,draw=black!75,fill=white,inner sep=0pt,
            minimum size=3.5pt,line width=.6pt},
  e/.style={line width=1.55pt,line cap=round},
  esmall/.style={line width=1.35pt,line cap=round},
  blank/.style={draw=black!40,densely dashed,line width=.85pt},
  ghost/.style={draw=black!12,line width=.6pt},
  subcard/.style={fill=black!2.5,draw=black!16,rounded corners=4pt,line width=.55pt},
  arr/.style={->,draw=black!75,line width=.8pt},
  lab/.style={font=\scriptsize,inner sep=1.5pt}
]
% Panel 1: H with deletion set S highlighted.
\begin{scope}[xshift=0cm]
  \foreach \i in {0,...,4} {\coordinate (a\i) at ({90+72*\i}:1.05);}
  \foreach \i/\j in {0/1,1/2,2/3,3/4,4/0} {\draw[blank] (a\i)--(a\j);}
  \foreach \i in {0,...,4} {\draw[blank] (0,0)--(a\i);}
  \fill[ecD!15] (0,0) circle[radius=.22];
  \node[v,draw=ecD,line width=.85pt] at (0,0) {};
  \node[lab,text=ecD!80!black] at (54:0.36) {$S$};
  \foreach \i in {0,...,4} {\node[v] at (a\i) {};}
  \node at (0,-1.60) {$H$};
  \node[lab,text=black!65] at (0,-1.96) {$\Delta \le D$};
\end{scope}

% Fork arrows: delete S and Euler-split H-S.
\draw[arr] (1.35,0.22) to[out=25,in=180] (2.95,1.18);
\draw[arr] (1.35,-0.22) to[out=-25,in=180] (2.95,-1.18);
\node[lab,align=center] at (2.10,0) {$-S$,\\Euler split};

% Panel 2 (stacked): parallel recursive colorings of H_1 and H_2.
\begin{scope}[xshift=4.05cm,yshift=1.18cm]
  \draw[subcard] (-0.92,-0.80) rectangle (2.18,0.88);
  \foreach \i in {0,...,4} {\coordinate (b\i) at ({90+72*\i}:0.66);}
  \foreach \i/\j in {0/1,1/2,2/3,3/4,4/0} {\draw[ghost] (b\i)--(b\j);}
  \draw[esmall,ecA] (b0)--(b1);
  \draw[esmall,ecA] (b2)--(b3);
  \draw[esmall,ecC] (b4)--(b0);
  \foreach \i in {0,...,4} {\node[vsmall] at (b\i) {};}
  \node[lab,align=left] at (1.42,0.06) {$H_1$ on $\mathcal{P}_1$};
\end{scope}
\begin{scope}[xshift=4.05cm,yshift=-1.18cm]
  \draw[subcard] (-0.92,-0.80) rectangle (2.18,0.88);
  \foreach \i in {0,...,4} {\coordinate (c\i) at ({90+72*\i}:0.66);}
  \foreach \i/\j in {0/1,1/2,2/3,3/4,4/0} {\draw[ghost] (c\i)--(c\j);}
  \draw[esmall,ecB] (c1)--(c2);
  \draw[esmall,ecB] (c3)--(c4);
  \foreach \i in {0,...,4} {\node[vsmall] at (c\i) {};}
  \node[lab,align=left] at (1.42,0.06) {$H_2$ on $\mathcal{P}_2$};
\end{scope}
\node[lab,text=black!65] at (4.68,-2.26) {$\mathcal{P}_1\cap\mathcal{P}_2=\varnothing$};

% Merge arrows: immediate union.
\draw[arr] (6.35,1.18) to[out=0,in=155] (7.55,0.22);
\draw[arr] (6.35,-1.18) to[out=0,in=-155] (7.55,-0.22);
\node[lab,align=center] at (6.95,0) {union};

% Panel 3: combined coloring on H-S.
\begin{scope}[xshift=8.90cm]
  \foreach \i in {0,...,4} {\coordinate (d\i) at ({90+72*\i}:1.05);}
  \foreach \i/\j/\col in {0/1/ecA,1/2/ecB,2/3/ecA,3/4/ecB,4/0/ecC}
    {\draw[e,\col] (d\i)--(d\j);}
  \foreach \i in {0,...,4} {\node[v] at (d\i) {};}
  \node at (0,-1.60) {$H-S$};
  \node[lab,text=black!65] at (0,-1.96) {$\Delta \le D-2$};
\end{scope}

% Arrow: star restoration.
\draw[arr] (10.25,0)--node[lab,above=3pt] {$+S$} node[lab,below=3pt,text=black!65] {Alg.~\ref{alg:restore}} (11.55,0);

% Panel 4: restored coloring of H in full palette P.
\begin{scope}[xshift=12.90cm]
  \foreach \i in {0,...,4} {\coordinate (f\i) at ({90+72*\i}:1.05);}
  \foreach \i/\j/\col in {0/1/ecA,1/2/ecB,2/3/ecA,3/4/ecB,4/0/ecC}
    {\draw[e,\col] (f\i)--(f\j);}
  \foreach \i/\col in {0/ecB,1/ecC,2/ecD,3/ecE,4/ecA}
    {\draw[e,\col] (0,0)--(f\i);}
  \node[v] at (0,0) {};
  \node[lab] at (54:0.38) {$S$};
  \foreach \i in {0,...,4} {\node[v] at (f\i) {};}
  \node at (0,-1.60) {$H$ in $\mathcal{P}$};
  \node[lab,text=black!65] at (0,-1.96) {$|\mathcal{P}|=D+1$};
\end{scope}
\end{tikzpicture}
\caption{One recursive level of Algorithm~\ref{alg:color}. Deleting $S$
lowers the maximum degree by at least two, so Euler splitting divides $H-S$
into two subgraphs colored recursively on disjoint subpalettes
$\mathcal{P}_1,\mathcal{P}_2\subseteq\mathcal{P}$. Their union colors $H-S$
immediately, and Algorithm~\ref{alg:restore} restores $S$ in the full palette
$\mathcal{P}$ (either one vertex at a time or in independent classes).}
\label{fig:recursion}
\end{figure}
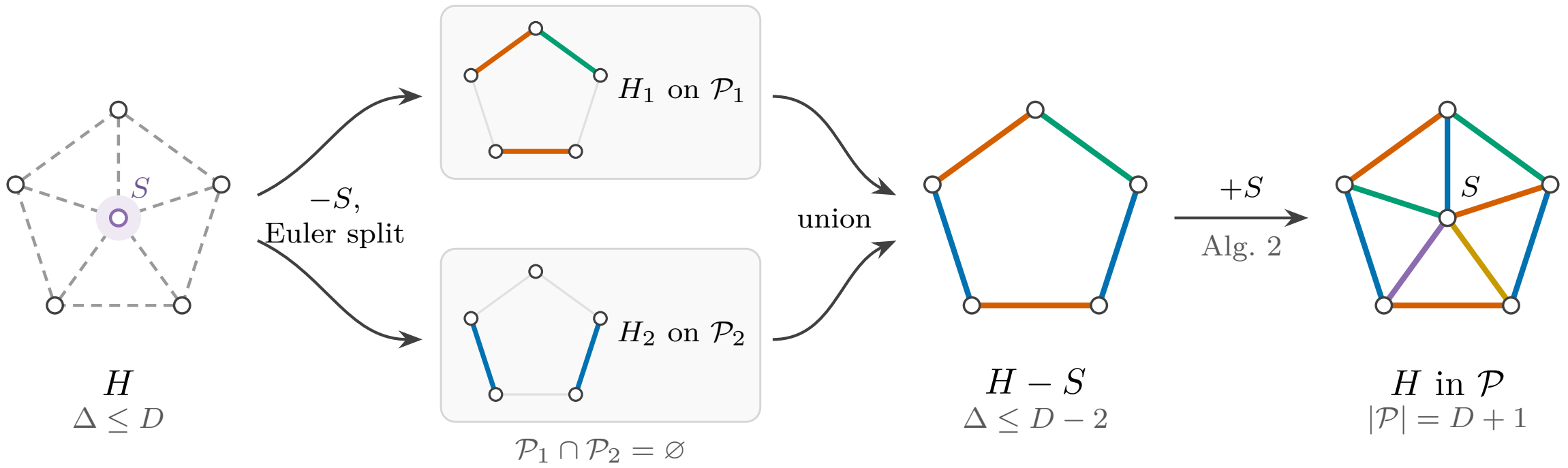

The drop of two has a specific purpose. We alternate red and blue around
Euler tours of the remaining graph, adding temporary edges to make degrees
even. A tour of odd length can give its starting vertex an extra red edge.
Accounting for this rounding and for one extra color in each recursive
call, the two palettes together need at most the remaining maximum degree
plus three colors. Lowering the maximum degree by two makes both palettes
fit in the parent palette, as shown in Lemma~\ref{lem:budget}. The two calls proceed
in parallel and their colorings can be combined immediately
(Figure~\ref{fig:recursion}).

Finally, we reinsert the removed vertices in one of two ways. For large
$D$, we restore them one by one for $\softO(n/D)$ depth at this level. For smaller $D$, we use the
neighborhood bound a second time: the subgraph induced by the removed
vertices has polylogarithmic maximum degree, so it splits into
polylogarithmically many independent sets with small overlap in the rest
of the graph. Restoring these sets one after another takes $\softO(D)$
depth. Choosing the faster mode gives
$\softO(\min\{D,n/D\})\le\softO(\min\{\Delta,\sqrt{n}\})$ depth at this
level. Because the
maximum degree halves at each split, there are only logarithmically many
levels, giving $\softO(\min\{\Delta,\sqrt{n}\})$ depth overall.

\subsection{Comparison with prior work}\label{sec:ov-compare}
\paragraph{Where the factor of $\Delta$ is saved.}
In Fineman and Mohammadi~\cite{FM26}, each recursive merge erases an
extra color class and repairs its uncolored edges. Although those edges
form a matching, many fan centers can still appear near the same
vertex. As a result, their randomized algorithm pays two separate
factors of $\Delta$ per round to isolate non-interfering fans. First,
it samples a matching of color pairs, activating each fan's requested
pair with probability $\Theta(1/\Delta)$. Second, it samples again at
rate $\Theta(1/\Delta)$ to avoid the $O(\Delta)$ other fans that can
still block a typical active fan, leaving fans whose rotations and
precomputed alternating paths can run together. Together, these two
steps color an $\Omega(1/\Delta^2)$ fraction of the uncolored edges in
expectation per round. Deterministically, their schedule resolves a
conflict graph of average degree $O(\Delta^2)$, giving an
$\softO(\Delta^3)$ depth bound. 

Our recursion instead chooses \emph{which vertices to
restore} before splitting. At a call of maximum degree $D$, every vertex
outside a restoration group has only polylogarithmically many neighbors in it
(Lemma~\ref{lem:delete}), so each fan shares vertices with only
$\softO(D)$ other fans. Conditional on a given fan being active,
only polylogarithmically many overlapping fans are active in expectation:
\begin{itemize}
\item \emph{Different color pairs.} Each overlapping fan asking for a
  different pair is active with probability at most $O(1/\Delta)$, so
  only polylogarithmically many of them are active on average.
\item \emph{Same color pair.} Fans asking for the same pair become
  active together. Because only $\softO(\Delta)$ centers share a
  neighbor with any given center, balancing center marks among them
  using Fineman and Mohammadi's bit-by-bit color exchanges ensures that
  only polylogarithmically many overlapping fans ask for the same pair.
\end{itemize}
Finally, we do not need an extra sampling step to keep alternating
paths apart. Instead of finding paths in advance, we rotate the chosen
disjoint fans first and resolve the two-color components afterward,
which lets a constant fraction of the rotated fans finish together.
Choosing a matching of color pairs is thus the only loss of order
$\Delta$: each phase deterministically colors an
$\Omega(1/(\Delta\log^{O(1)} n))$ fraction of the uncolored edges
(Lemma~\ref{lem:joint}), saving one factor of $\Delta$ over Fineman and
Mohammadi's randomized depth bound and two over their deterministic
bound.

\paragraph{Relation to separable fans.}
Postponing the finishing paths works because every fan that needs one
keeps two ways to stop. When two walks meet, they give two uncolored
spokes whose leaves miss the same color, as in the separable fans of
Assadi et al.~\cite{ABBCSZ26R}. Here a single uncolored spoke can also
give two stopping points when its fan runs into a cycle; choosing
between them lets us postpone the finishing paths until after the other
fans have rotated.

\paragraph{Work and processors.}
Khuzman and Elkin~\cite{EK26} process fans requesting a common color
pair. Fineman and Mohammadi process many disjoint pairs in one round.
Our algorithm uses more processors to evaluate many possible rotations
on separate copies. Table~\ref{tab:resources} compares their guarantees with ours across depth and number of processors.
\begin{table}[ht]
\centering\small
\setlength{\tabcolsep}{7pt}
\begin{tabular}{@{}lcc@{}}
\hline
Algorithm & Depth & Processors\\
\hline
Khuzman--Elkin, det. & $\Delta^4$ & $m\Delta$\\
Fineman--Mohammadi, det. & $\Delta^3$ & $m$\\
Fineman--Mohammadi, rand. & $\Delta^2$ & $m$\\
This paper, det. & $\min\{\Delta,\sqrt n\}$ & $n^3\Delta$\\
\hline
\end{tabular}
\caption{Depth and processors bounds up to polylogarithmic factors.}
\label{tab:resources}
\end{table}

\section{Preliminaries}\label{sec:prelim}
The input is an edge list for a simple undirected graph on vertices with
distinct identifiers in $[n]$. We omit an empty edge list from a recursive
call and assign maximum degree zero to an empty graph. The original $n$
and $L=\ceil{\log_2(n+2)}$ are fixed throughout. In asymptotic bounds we
take logarithms to be at least one. A subproblem uses a
palette of $q$ labels, locally numbered $0,\ldots,q-1$ when convenient.
We write $N(v)$ for the neighbors of $v$ in the graph of the current call.

A partial coloring is proper on the edges already colored. We call an
uncolored edge a \emph{blank}. A color is \emph{missing} at a vertex when
none of its incident edges has that color. In a graph of maximum degree
at most $q-1$, a vertex incident to $d$ blanks misses at least $d+1$
colors. For distinct colors $a,b$, the spanning subgraph of edges colored
$a$ or $b$ consists of alternating paths, even cycles, and isolated vertices.
Exchanging $a,b$ on a whole component preserves properness.
Exchanges on different components of one pair, or on disjoint pairs
of colors, can be applied simultaneously.

\paragraph{Parallel operations.}
We use a uniform CREW PRAM with $O(L)$-bit words. Read-only data can be
shared. Simultaneous trials start from the same coloring, use separate
output cells, and commit only the selected trial. Ties are broken by the
least identifier or seed. A uniform seed in a proof denotes an average
over an explicitly enumerable list. Every algorithm we execute is
deterministic.

\begin{lemma}[Parallel primitives, Batcher~\cite{Batcher68} and Atallah--Vishkin~\cite{AV84}]\label{lem:pram}
A list of $t$ word-size records, with $t$ polynomial in $n$, can be sorted
deterministically on CREW in $O(\log^2 t)$ depth with $O(t)$
processors. On a connected undirected even-degree graph with $v$ vertices
and $e>0$ edges, a cyclic Euler tour can be found in $O(L^3)$
CREW depth with $O(v+e)$ processors.
\end{lemma}
\begin{proof}
Batcher's sorting network has $O(\log^2 t)$ comparator layers, each on
disjoint pairs of records. Atallah and Vishkin give a logarithmic-time
CRCW Euler-tour algorithm with a linear number of processors. We simulate
each of its $O(L)$ steps in $O(L^2)$ depth by sorting write requests
by address and writer, then committing the first request per address. Cutting the cyclic tour
at a designated incidence and ranking the resulting list by pointer doubling
lets us start at any specified vertex.
\end{proof}
Prefix sums, minima, and pointer doubling take $O(L)$ depth with one
processor per record. To identify components in a $v$-vertex graph, we
square its reflexive Boolean adjacency matrix $O(L)$ times. Each entry is
an OR reduction of depth $O(L)$, giving $O(L^2)$ total depth with
$O(v^3)$ processors. For a two-color subgraph, $O(n)$ processors suffice.
Each directed incidence follows the other edge at its head until the path
ends. Doubling these pointers in both directions returns the component's
least vertex identifier and, on a path, the endpoint in each direction.
An isolated vertex supplies its own identifier.

\paragraph{Finite choices.}
On $t$ distinct binary identifiers of $\ceil{\log_2(t+1)}$ bits, all
affine binary forms $x\mapsto \langle u,x\rangle_2+v$ give fair,
pairwise independent bits with $O(t)$ seeds. The subscript denotes the
binary dot product, and addition here is modulo two. Over a prime field
$\F_P$, the $P^2$ maps $x\mapsto ux+v$ with $(u,v)\in\F_P^2$ give
pairwise independent uniform residues on distinct inputs. Counting residues below an
integer threshold realizes its exact fraction of $P$. We choose the least
prime in a specified interval $[b,2b]$ for integer $b\ge2$. Bertrand's
postulate guarantees existence and parallel trial division finds it. For
word-size polynomially bounded $b$, the division of two word-size integers
can itself be done in $O(L^2)$ elementary steps. The program generates
its seeds and field representations from these sizes.

\paragraph{Balancing many counts at once.}
Both vertex deletion and the distribution of missing colors require a
deterministic substitute for halving. Lemma~\ref{lem:sign} constructs signs
for a sparse integer matrix with entries of absolute value at most two.
Every row $h$ simultaneously satisfies
\begin{equation}\label{eq:sign}
           |\langle h,\sigma\rangle|
           \le \|h\|_1/(8L)+O(L^3).
\end{equation}
For $R$ rows, $s$ columns, and $M$ nonzeros, all polynomial in $n$,
the depth is $O(L^{12})$ using $O(s(M+R+s))$ processors for $s\ge1$. The power
of $L$ is a conservative bound for the word implementation.

We use the following consequence in both applications. If a count $U$ is
split into two children $(U\pm\langle h,\sigma\rangle)/2$ with
$\|h\|_1\le U$, either child lies between
$(1/2-1/(16L))U-O(L^3)$ and $(1/2+1/(16L))U+O(L^3)$.
At each descendant the row is formed from the items contributing to
its current count $U$, so $\|h\|_1\le U$ holds again. Starting from
a count $u$, any descendant after $j\le L$ splits therefore
has count between
\begin{equation}\label{eq:halving}
                    c\,u\,2^{-j}-C L^3
           \quad\text{and}\quad C\,u\,2^{-j}+C L^3
\end{equation}
for absolute $c,C>0$. Indeed, the $j$th powers of both multipliers are
within constant factors of $2^{-j}$, and the additive errors sum
geometrically.

\section{The complete coloring algorithm}\label{sec:global}
A recursive call receives a graph $H$ of maximum degree $D$ and an
interval $\mathcal P$ of $D+1$ colors. It returns a complete coloring
using that interval. Its only coloring subroutine, Algorithm~\ref{alg:restore},
takes a proper partial coloring whose blanks are all incident to an
independent set $I$. It may change colors throughout the current graph.
It fills the blanks in polylogarithmic depth for one vertex, and in
$\softO(q(1+\tau))$ depth when each outside vertex has at most $\tau$
neighbors in $I$. Lemma~\ref{lem:restore} gives its precise contract.

\subsection{Algorithm}
Algorithm~\ref{alg:color} summarizes one recursive level (illustrated in
Figure~\ref{fig:recursion}).
\begin{algorithm}[H]
\caption{Recursive edge coloring}\label{alg:color}
\io{A graph $H$ of maximum degree $D$ and an interval $\mathcal P$ of $D+1$ colors.}
   {A proper coloring of $E(H)$ with colors in $\mathcal P$.}
\begin{steps}
\st{If $D\le1$, color every edge with the first label of $\mathcal P$ and return.}
\st{Form the nested sets below. Keep the last $S$ with $\Delta(H-S)\le D-2$.}
\st{Split the edges of $H-S$ into red and blue by the Euler tours below.}
\st{For each nonempty part $i\in\{1,2\}$, allocate a disjoint subinterval of $D_i+1$ labels, where $D_i$ is its maximum degree. Color the parts recursively in parallel.}
\st{Combine their edge colorings on $H-S$.}
\st{If $D^2\le n$, partition $S$ into independent classes as below. Otherwise use one singleton class per vertex.}
\st{For each class $I$, in order:}
\st[1]{Add $I$ and all edges between $I$ and the vertices already present. Leave these edges blank. Run Algorithm~\ref{alg:restore} with $I$ and the full palette $\mathcal P$.}
\st{Return the resulting coloring of $H$.}
\end{steps}
\end{algorithm}

\begin{itemize}[leftmargin=*,itemsep=4pt]
\item \textbf{Vertex deletion.} We start with $S_0=V(H)$ and form
$S_1,\ldots,S_L$. In each step we apply~\eqref{eq:sign} to the indicator
rows for all neighborhoods and for $V(H)$, restricted to the current set.
Vertices with a plus sign form the next set. Since $S_0$ passes the degree
test, the algorithm always selects a set.
\item \textbf{Euler splitting.}  If $H-S$ has any odd-degree vertices, we add
one auxiliary vertex adjacent to all of them. All degrees are now even. In each
nontrivial component we compute an Euler tour using Lemma~\ref{lem:pram}
and alternate red and blue, starting red. If present, the auxiliary vertex
starts its component's tour. We then discard any auxiliary edges.
\item \textbf{Vertex reinsertion.} For the independent classes we set
$h=\Delta(H[S])$ and choose a prime $P=\Theta(n)$ with $P\ge4n$.
For each affine map on $\F_P$, we select the remaining vertices of $S$
whose identifier maps below $\floor{P/(2(h+1))}$. We keep each selected
vertex with no selected neighbor among the remaining vertices of $S$.
These vertices form an independent set. We remove the largest
set so obtained and repeat until $S$ is exhausted.
\end{itemize}

\subsection{Analysis}
The deletion set has to meet two requirements. Its small size makes
single-star insertion affordable. Its small intersection with every
neighborhood allows an independent class to be restored in one call.

\begin{lemma}\label{lem:delete}
For $D\ge2$, the chosen set $S$ satisfies $\Delta(H-S)\le D-2$,
$|S|=O(nL^3/D)$, and $\max_v|N(v)\cap S|=O(L^3)$.
\end{lemma}
\begin{proof}
For $D\ge A L^3$, choose $i\le L$ with
$A L^3\le D/2^i<2A L^3$. For a sufficiently large absolute $A$, the
lower bound in~\eqref{eq:halving} gives every vertex of degree at least
$D-1$ at least two neighbors in $S_i$. Thus $S_i$ passes the degree test.
The last passing set is contained in $S_i$ because the sets are nested.
Applying the upper bound
of~\eqref{eq:halving} to the whole vertex set and to each neighborhood
proves both size bounds. For $D<A L^3$, these bounds follow from $|S|\le n$
and $|N(v)\cap S|\le D$. In both cases the degree bound is the test itself.
\end{proof}

\begin{lemma}\label{lem:budget}
Let $D_0=\Delta(H-S)$ and let $D_1,D_2$ be the red and blue maximum degrees.
Then
\begin{equation}\label{eq:budget}
 D_1\le\floor{D_0/2}+1,\qquad D_2\le\ceil{D_0/2},\qquad
 (D_1+1)+(D_2+1)\le D_0+3\le D+1 .
\end{equation}
\end{lemma}
\begin{proof}
At a vertex other than the start of a tour, each arrival and departure
have opposite colors. In the auxiliary component every ordinary vertex
has at most $\ceil{D_0/2}$ edges of each color before its auxiliary edge,
if any, is removed. Removing that edge decreases one color degree by one.
In every other component all degrees are even. Only the start of an
odd-length tour has unequal counts, with one extra red and one fewer blue
edge than half its degree. This gives the two individual bounds. Adding them and using
Lemma~\ref{lem:delete} gives the palette inequality.
\end{proof}

\begin{proposition}[Recursive guarantee]\label{prop:recursive}
Given a subgraph with at least one edge, maximum degree $D$, and an
interval of $D+1$ colors, Algorithm~\ref{alg:color} returns a proper
edge coloring in $O(L^{24}\min\{D,\sqrt n\})$ depth with
$O(n^3D)$ processors.
\end{proposition}
\begin{proof}
Since $D_0\le D-2$ by Lemma~\ref{lem:delete}, the larger bound
$\floor{D_0/2}+1$ in~\eqref{eq:budget} is at most $\floor{D/2}$.
Induction supplies their colorings, which combine because their intervals
are disjoint. Before a class is restored, the graph on the previously
present vertices is completely colored. Every newly added edge joins
the independent class to an old vertex and starts blank. The maximum
degree stays at most $D$, and the full $D+1$ palette is available.
Lemma~\ref{lem:restore} applies, including when a restoration recolors
edges of both children. Each original edge is introduced exactly once,
when both its endpoints are present.

We bound the depth outside the recursive calls. In the class construction,
the marking rate $r$ is $\Theta(1/(h+1))$ with $rh\le1/2$. Among $t$
remaining vertices, the number of isolated marks is at least the number
marked minus twice the number of edges with both ends marked. Its
expectation over the maps is at least $rt-r^2ht\ge rt/2$. The best
class removes at least this many vertices, hence at least one whenever
$t>0$. Geometric decay exhausts $S$ in $O((h+1)L)=O(L^4)$ rounds. Each class has
overlap $O(L^3)$ by Lemma~\ref{lem:delete}. Lemma~\ref{lem:restore}
gives $O(DL^{20})$ depth per class, or $O(DL^{24})$ in total.
The singleton method takes $O(|S|L^{17})=O((n/D)L^{20})$ depth.
Vertex deletion takes $O(L^{13})$ depth. The Euler tours and each round
of affine-map trials take $O(L^3)$ depth, within these per-call bounds.

Each parent waits only for its own two children, so total depth is bounded by the
largest sum of local costs along a root-to-leaf path. Write $d$ for the
maximum degree of the subgraph at a node on this path. While $d>\sqrt n$, the quantities $n/d$ at
least double from one call to the next and are all at most $\sqrt n$.
Their sum is $O(\sqrt n)$, and this regime occurs only if $D>\sqrt n$.
Once $d\le\sqrt n$, successive degrees at least halve and sum to
$O(\min\{D,\sqrt n\})$. Multiplying these two sums by $L^{20}$ and
$L^{24}$, respectively, proves the depth bound.

A nonempty node of degree $d$ uses $O(n^3d)$ processors. The deletion matrix
has $O(n)$ rows and columns and $O(nd)$ nonzeros. Connectivity and
Euler tours fit by Lemma~\ref{lem:pram} and the Boolean computation in
Section~\ref{sec:prelim}. The $O(n^2)$ affine trials each inspect
$O(nd)$ incidences. The restoration cost is in Lemma~\ref{lem:restore}.
To allocate processors uniformly, we reserve $O(n^3)$ per color in a
node's interval. The children inherit disjoint blocks by~\eqref{eq:budget}.
Once both return, the parent reuses its whole block. Taking $H=G$ proves
Theorem~\ref{thm:main}.
\end{proof}

\section{Restoring uncolored stars}\label{sec:rest}
Let $I$ be an independent set incident to every blank. We call its vertices
\emph{centers} and their incident edges \emph{spokes}. Write $d_s$ for the
number of blank spokes at $s$, and $d=\sum_{s\in I}d_s$. Each blank is
counted once. The overlap $\tau=\max_{x\notin I}|N(x)\cap I|$ is zero if
the maximum is empty. We keep the graph, $I$, and the palette fixed
throughout a restoration. The coloring can change anywhere in this graph.

\subsection{Algorithm}
Each phase starts in the current coloring. We designate one missing color
at each vertex as its \emph{mark}. The procedure of Section~\ref{sec:fans}
recolors the graph while keeping all marks missing and all blanks in place.
It spreads marks over $N(s)$ and over the set of potential competing centers
\[
                 W_s=\{t\in I\setminus\{s\}:N(t)\cap N(s)\ne\varnothing\}.
\]
Let $B$ be the maximum multiplicity of a single mark in any $N(s)$,
and define $B'$ analogously for the sets $W_s$. The sets have sizes at most $q-1$
and $\tau(q-1)$, so Lemma~\ref{lem:spread} gives
\begin{equation}\label{eq:mult}
                       B=O(L^3),\qquad B'=O(\tau+L^3).
\end{equation}

Lemma~\ref{lem:fans} constructs three kinds of choices. An \emph{exit}
moves a blank to a leaf whose mark $b$ is also missing at the center.
A \emph{two-blank choice} rotates two paths to distinct leaves with the
same mark $b$. A \emph{nested fan} offers a short prefix ending at $y$
and a long prefix ending at $z$, both marked $b$. We call $b$ the
\emph{terminal mark} in all three cases. The leaf set includes every
neighbor traversed by the choice, through the full long prefix for a
nested fan. Distinct choices at one center have disjoint leaf sets.
For a nondirect choice with assigned missing center color $a$, a
terminal is \emph{usable} if it still misses $b$ and lies outside the
center's current $(a,b)$-component.

\begin{algorithm}[H]
\caption{Star restoration}\label{alg:restore}
\io{A proper partial coloring of a graph of degree at most $q-1$ on a palette of $q$ colors, and an independent set $I$ incident to every blank.}
   {A complete proper coloring on the same graph and palette.}
\begin{steps}
\st{While a blank remains:}
\st[1]{Mark the least missing color at each vertex. Spread marks on all $N(s),W_s$ and build the fan choices (Section~\ref{sec:fans}).}
\st[1]{If $I=\{s\}$:}
\st[2]{Keep one choice per terminal mark. Reserve the exit marks and assign distinct other missing colors $a$ to the nondirect choices.}
\st[2]{For each nested fan, test the short-prefix terminal $y$ in the current coloring. Take the long prefix if $y$ belongs to the $(a,b)$-component of $s$, and the short prefix otherwise.}
\st[2]{Rotate and fill the exits. Rotate both paths of each two-blank choice and the selected nested prefixes.}
\st[2]{For each nondirect choice, select a usable terminal in this coloring. Exchange its component and fill its blank with $a$, in parallel over the choices.}
\st[1]{Otherwise, label each choice by $(a_s,b)$, with $a_s$ the center's mark. For an exit with $b=a_s$, use the least color other than $a_s$ as the second scheduling label.}
\st[2]{Keep one choice per ordered pair at each center. Commit the coloring from Algorithm~\ref{alg:joint}.}
\st{Return the coloring.}
\end{steps}
\end{algorithm}

An exit fills its blank with its terminal mark $b$, including when its
scheduling label differs. At several centers a nondirect choice may fill
its designated blank with either $a_s$ or $b$. A two-blank choice
designates at most one of its vacated spokes. Every unfilled spoke stays
blank at the same center, and the next phase rebuilds its marks and fans.

\subsection{Analysis}
For one center the essential issue is competition for colors at that
vertex. With several centers we keep only one choice at each active center
and control the leaves they share. The two cases have different sources
of progress.

\begin{lemma}[Star restoration]\label{lem:restore}
For $2\le q\le n+1$, Algorithm~\ref{alg:restore} completes every input
in its contract, with $O(n^3q)$ processors. At one center it takes
$O(L^4)$ phases and $O(L^{17})$ depth. For arbitrary $I$ it takes
$O(qL^4(\tau+L^3))$ phases and $O(qL^{17}(\tau+L^3))$ depth.
After every phase the coloring is proper and each remaining blank
is incident to its original center.
\end{lemma}
\begin{proof}
At one center, Lemma~\ref{lem:fans} gives at least $d_s/3$ choices.
Every terminal mark occurs at most $B$ times among them, since the leaf
sets are disjoint. At least $d_s/(3B)$ choices survive. Each choice uses
at least one original blank, so there are at most $d_s$ choices. The center
has at least $d_s+1$ missing colors. After reserving the distinct exit
marks, it therefore has enough missing colors to give every remaining
choice its own $a$. For those choices $b$ is present at $s$. Thus all
pairs $\{a,b\}$ are disjoint and avoid every exit reservation.

For a nested fan, the spoke of color $b$ belongs to its long prefix. No
other choice touches it or introduces the missing color $a$. Hence the
other rotations and exit fills at this center leave the entire
$(a,b)$-subgraph unchanged. The short prefix also leaves this subgraph
unchanged. The test on the original component in Algorithm~\ref{alg:restore}
therefore chooses a usable terminal by Lemma~\ref{lem:terminal}. For a
two-blank choice, another rotation may move the center's $b$-spoke.
We test both terminals \emph{after} the rotations. Both still miss $b$
and the center still misses its reserved $a$, so the lemma again applies.
Swapping the terminal component leaves $a$ missing at
the center and makes it missing at that terminal. All such exchanges and
fills are compatible, including at the common center, by disjointness of
the color pairs. Each phase fills at least $d_s/(3B)$ blanks.

At several centers there are initially at least $d/3$ choices. Before
keeping one per ordered pair, a pair occurs at most $2B$ times at a given
center. One terminal mark gives at most $B$ choices, and only the exit mark
$a_s$ can be reassigned to a different scheduling label. Thus $\Omega(d/B)$
choices enter Algorithm~\ref{alg:joint}. Lemma~\ref{lem:joint} shows that
the number of blanks decreases by at least
\[
                      \frac{c\,d}{qB(1+\tau+B')}
\]
for an absolute $c>0$. Its rotations move blanks only along spokes of their
original centers, and all exchanges preserve properness.

Both arguments hold from every current state. By~\eqref{eq:mult} and
$d\le n^2$, multiplicative decrease gives the stated phase counts, including
the last phase since the count is integral. Spreading uses $O(L)$ signing
stages of $O(L^{12})$ depth. Fan construction and the joint choices fit
within the resulting $O(L^{13})$ depth per phase by Sections~\ref{sec:fans}
and~\ref{sec:choices}. The single-center component tests and exchanges use
only disjoint color pairs. All these procedures use $O(n^3q)$ processors,
reused each phase. This proves both depth bounds.
\end{proof}
The remaining sections give the constructions used in a phase. We first
explain how the disjoint alternatives arise at one center. We then prove
that the choices made at distinct centers leave enough usable terminals.

\section{Marks and local fan rotations}\label{sec:fans}
The input is a proper partial $q$-coloring with one missing color marked
at each vertex. We first spread the marks over a prescribed family of
$O(n)$ sets. The resulting coloring is the common starting point for
the fan choices and the single-center component tests.

\subsection{Algorithm}
\paragraph{Spreading marks.}
We use the bitwise color pairing of Fineman and Mohammadi~\cite{FM26}.
On a subpalette of $Q=2^j$ colors we index its labels by $j$ bits.
At stage $i$ we pair labels that differ only in bit $i$. On each
two-color component $C$ of each pair, including isolates, we choose one
sign $\sigma_C$ that either retains or exchanges the two labels. We
apply the same permutation to any mark of those colors at its vertices.

A row corresponds to a prescribed set $F$ and a prefix of $i-1$ bits.
Among marks in $F$ belonging to that prefix in the subpalette, let $U$
be their number. For a component $C$, let $h_C$ be the number of those
marks in $C$ of its paired colors with next bit zero minus the number
with next bit one. When the plus sign retains the labels, the counts in
the two child prefixes are $(U\pm\sum_C h_C\sigma_C)/2$. We apply the
signing routine of Section~\ref{sec:balance} to all rows at once. For a
general $q$ we use the first and last $2^{\floor{\log_2 q}}$ labels as
two subpalettes, in that order. We use a single pass if they coincide.

\paragraph{Constructing the fans.}
At a center $s$, a neighbor $x$ points to the unique $y$ such that $sy$
carries the mark at $x$. If no such $y$ exists, $x$ is an exit. A blank
neighbor has indegree zero. A simple successor path $x_0,\ldots,x_h$
starting at a blank encodes a rotation. We move the old color of $sx_{i+1}$
to $sx_i$ for $0\le i<h$ and leave $sx_h$ blank. A path of length zero
leaves its starting spoke blank.

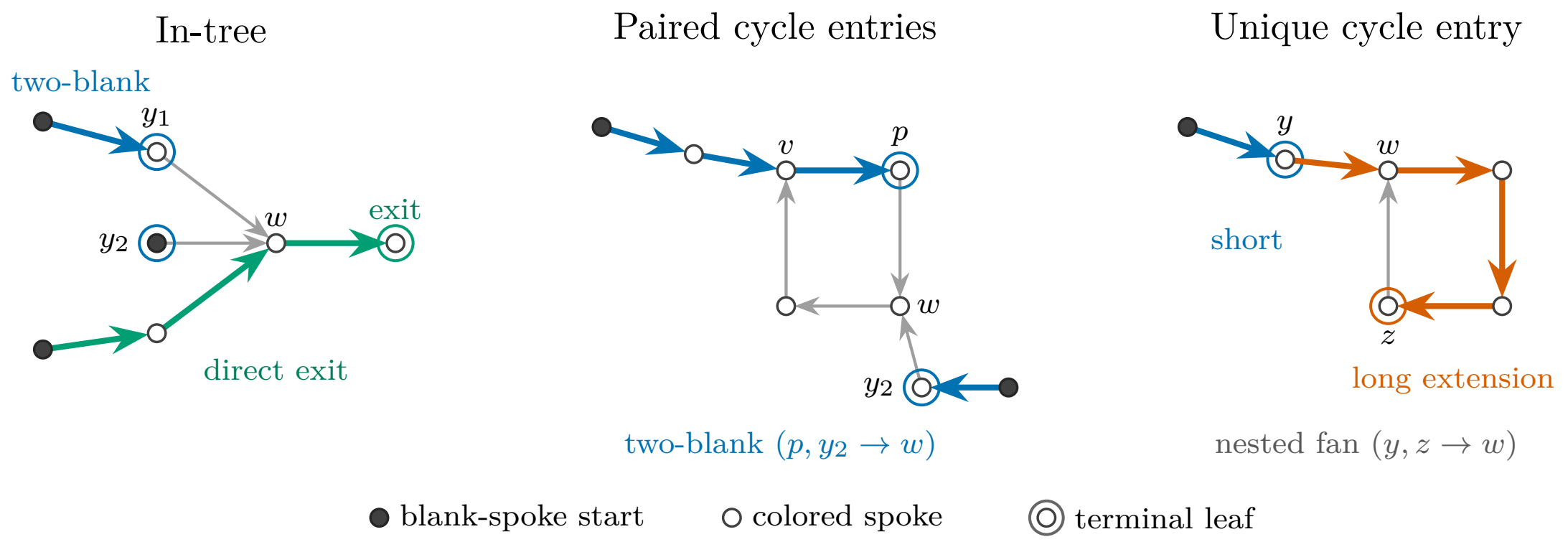
\begin{figure}[tbp]
\centering
% Functional graph on N(s) and token pairing (Section 6.1, Lemma lem:fans).
% Filled nodes denote blank-spoke neighbors (indegree-0 token starts).
% Three panels show: (1) in-tree yielding a two-blank choice + an exit;
% (2) cycle with two occupied entries yielding a two-blank choice;
% (3) cycle with one occupied entry yielding a nested fan (short y, long z).
% Requires TikZ arrows.meta, calc, and colors ecA, ecB, ecC.
\begin{tikzpicture}[
  x=1cm,y=1cm,font=\small,>=Stealth,
  v/.style={circle,draw=black!75,fill=white,inner sep=0pt,
            minimum size=4.6pt,line width=.65pt},
  start/.style={circle,draw=black!85,fill=black!75,inner sep=0pt,
            minimum size=4.6pt,line width=.65pt},
  ptr/.style={->,draw=black!38,line width=.8pt,shorten >=2.2pt,shorten <=2pt},
  pathB/.style={->,draw=ecB,line width=1.55pt,shorten >=2.2pt,shorten <=2pt},
  pathC/.style={->,draw=ecC,line width=1.55pt,shorten >=2.2pt,shorten <=2pt},
  pathA/.style={->,draw=ecA,line width=1.55pt,shorten >=2.2pt,shorten <=2pt},
  lab/.style={font=\scriptsize,inner sep=1.2pt}
]
% Panel 1: In-tree rooted at an exit.
\begin{scope}[xshift=0cm]
  \node[font=\small] at (1.55,2.15) {In-tree};
  \coordinate (u1) at (0,1.30);
  \coordinate (v1) at (1.05,1.02);
  \coordinate (u2) at (1.05,0.18);
  \coordinate (u3) at (0,-0.80);
  \coordinate (v3) at (1.05,-0.65);
  \coordinate (w0) at (2.15,0.18);
  \coordinate (ex) at (3.25,0.18);

  % Base pointers (untraversed edges).
  \draw[ptr] (v1)--(w0);
  \draw[ptr] (u2)--(w0);

  % Two-blank choice stopping before w0 (blue).
  \draw[pathB] (u1)--(v1);
  \draw[ecB,line width=.75pt] (v1) circle[radius=4.6pt];
  \draw[ecB,line width=.75pt] (u2) circle[radius=4.6pt];

  % Surviving third token reaching exit (green).
  \draw[pathC] (u3)--(v3);
  \draw[pathC] (v3)--(w0);
  \draw[pathC] (w0)--(ex);
  \draw[ecC,line width=.75pt] (ex) circle[radius=4.6pt];

  \foreach \p in {u1,u2,u3} {\node[start] at (\p) {};}
  \foreach \p in {v1,v3,w0,ex} {\node[v] at (\p) {};}

  \node[lab,above=5pt] at (v1) {$y_1$};
  \node[lab,left=5.5pt] at (u2) {$y_2$};
  \node[lab,above=3pt] at (w0) {$w$};
  \node[lab,above=5pt,text=ecC!80!black] at (ex) {exit};
  \node[lab,text=ecB] at (0.35,1.68) {two-blank};
  \node[lab,text=ecC!80!black] at (2.15,-0.98) {direct exit};
\end{scope}

% Panel 2: Cycle with two occupied entries.
\begin{scope}[xshift=5.15cm]
  \node[font=\small] at (1.60,2.15) {Paired cycle entries};
  \coordinate (t1) at (0.00,1.25);
  \coordinate (y1) at (0.85,1.00);
  \coordinate (cv) at (1.70,0.85);
  \coordinate (cp) at (2.75,0.85);
  \coordinate (cw) at (2.75,-0.40);
  \coordinate (cq) at (1.70,-0.40);
  \coordinate (t2) at (3.75,-1.15);
  \coordinate (y2) at (2.95,-1.15);

  % Unused cycle pointers.
  \draw[ptr] (cp)--(cw);
  \draw[ptr] (cw)--(cq);
  \draw[ptr] (cq)--(cv);
  \draw[ptr] (y2)--(cw);

  % First token extends through v to cycle predecessor p of w;
  % second token stays at y2 just before w.
  \draw[pathB] (t1)--(y1);
  \draw[pathB] (y1)--(cv);
  \draw[pathB] (cv)--(cp);
  \draw[pathB] (t2)--(y2);
  \draw[ecB,line width=.75pt] (cp) circle[radius=4.6pt];
  \draw[ecB,line width=.75pt] (y2) circle[radius=4.6pt];

  \foreach \p in {t1,t2} {\node[start] at (\p) {};}
  \foreach \p in {y1,cv,cp,cw,cq,y2} {\node[v] at (\p) {};}

  \node[lab,above=3pt] at (cv) {$v$};
  \node[lab,above=5pt] at (cp) {$p$};
  \node[lab,right=3pt] at (cw) {$w$};
  \node[lab,left=5.5pt] at (y2) {$y_2$};
  \node[lab,text=ecB] at (1.65,-1.68) {two-blank ($p, y_2 \to w$)};
\end{scope}

% Panel 3: Cycle with one occupied entry -> nested fan.
\begin{scope}[xshift=10.55cm]
  \node[font=\small] at (1.65,2.15) {Unique cycle entry};
  \coordinate (s0) at (0.00,1.25);
  \coordinate (yy) at (0.90,0.95);
  \coordinate (ww) at (1.85,0.85);
  \coordinate (c1) at (2.90,0.85);
  \coordinate (c2) at (2.90,-0.40);
  \coordinate (zz) at (1.85,-0.40);

  % Closing cycle pointer z -> w (unused by long prefix).
  \draw[ptr] (zz)--(ww);

  % Short prefix: s0 -> y (blue).
  \draw[pathB] (s0)--(yy);
  \draw[ecB,line width=.75pt] (yy) circle[radius=4.6pt];

  % Long extension: y -> w -> c1 -> c2 -> z (vermilion).
  \draw[pathA] (yy)--(ww);
  \draw[pathA] (ww)--(c1);
  \draw[pathA] (c1)--(c2);
  \draw[pathA] (c2)--(zz);
  \draw[ecA,line width=.75pt] (zz) circle[radius=4.6pt];

  \node[start] at (s0) {};
  \foreach \p in {yy,ww,c1,c2,zz} {\node[v] at (\p) {};}

  \node[lab,above=5pt] at (yy) {$y$};
  \node[lab,above=3pt] at (ww) {$w$};
  \node[lab,below=5pt] at (zz) {$z$};
  \node[lab,text=ecB] at (0.55,0.22) {short};
  \node[lab,text=ecA] at (2.45,-1.08) {long extension};
  \node[lab,text=black!65] at (1.65,-1.68) {nested fan ($y, z \to w$)};
\end{scope}

% Compact legend at bottom.
\begin{scope}[yshift=-2.35cm,xshift=3.1cm]
  \node[start] at (0,0) {};
  \node[lab,right=4pt] at (0,0) {blank-spoke start};
  \node[v] at (3.25,0) {};
  \node[lab,right=4pt] at (3.25,0) {colored spoke};
  \draw[black!60,line width=.75pt] (6.15,0) circle[radius=4.4pt];
  \node[v] at (6.15,0) {};
  \node[lab,right=6pt] at (6.15,0) {terminal leaf};
\end{scope}
\end{tikzpicture}
\caption{Constructing disjoint fan choices on the pointer graph of $N(s)$.
Tokens start at blank-spoke neighbors (filled circles) and follow each leaf's
mark pointer. Left: in an in-tree, paired arrivals before $w$ form a
two-blank choice ($y_1,y_2$) while an unpaired token reaches an exit.
Center: two occupied cycle entries pair so the first extends through $v$ to
the cycle predecessor $p$ of $w$ and the second stops at $y_2$ before $w$.
Right: a single waiting token before $w$ produces a nested fan with short
leaf $y$ and long leaf $z$.}
\label{fig:token-fans}
\end{figure}

We start a token at each blank neighbor, which has no predecessors, and
send it along the pointers. In an in-tree,
at a vertex $w$ we pair arrivals from its children in identifier order.
Both tokens in a pair stop at the children they came from. Only an
unpaired token enters $w$. At an exit that token produces a direct choice.
At a vertex on a directed cycle we first pair off-tree arrivals in the
same manner, leaving the unpaired token, if any, just before the entry.
We order the occupied entries cyclically starting at the least one.
If there are at least two, we pair consecutive entries, discarding the
last token when their number is odd. For a pair of entries $v,w$, the
first token extends through $v$ to the cycle predecessor of $w$. The
second stays just before $w$. If exactly one token remains on the whole
cycle after the off-tree pairings, waiting before $w$, we record both its
prefix ending at its current leaf $y$ and its extension through $w$ to the
cycle predecessor $z$ of $w$ (Figure~\ref{fig:token-fans}).

\subsection{Analysis}
\begin{lemma}[Spreading missing colors]\label{lem:spread}
For any $O(n)$ prescribed vertex sets, the procedure above keeps all
marks missing and every blank in place. In the resulting coloring every
color occurs as a mark $O(|F|/q+L^3)$ times on each set $F$. The depth
is $O(L^{13})$ with $O(n^3q)$ processors.
\end{lemma}
\begin{proof}
A whole-component exchange preserves properness and transports each
affected missing color to another missing color, even at an isolate.
A path can carry relevant marks only at its two endpoints. A cycle
carries none and an isolate carries at most one. Thus $|h_C|\le2$ and
$\sum_C|h_C|\le U$. The upper bound of~\eqref{eq:halving} after $j$ stages
gives $O(|F|/Q+L^3)$ at every label in the subpalette. The second pass
starts with at most $|F|$ marks in its subpalette and leaves every label
outside it undisturbed. Since $Q>q/2$, this proves the general bound.

There are $O(nq)$ rows and component signs per stage. Each set--vertex
incidence contributes to at most one entry at that stage. Summing sparse
records with the same row and column gives $O(n^2)$ nonzeros. The signing
bound in Section~\ref{sec:prelim} and $q\le n+1$ give $O(n^3q)$ processors.
There are $O(L)$ stages. We form the sets $W_s$ by a Boolean matrix product
on center--neighbor incidences within the same processor budget.
\end{proof}

\begin{lemma}[Choices at a center]\label{lem:fans}
From $d_s$ blank spokes the token construction produces at least $d_s/3$
choices with disjoint leaf sets. Each is an exit, two paths from different
blanks to distinct leaves marked with one color $b$, or the nested fan
from one blank to distinct leaves $y,z$ marked $b$. In the nested case
the short prefix leaves the $b$-spoke $sw$ in place and the long prefix
moves its color to $sy$. The construction uses $O(q^3)$ processors per
center and $O(L^2)$ depth.
\end{lemma}
\begin{proof}
Each arriving color in a rotation is missing at its receiving leaf, and
the center keeps precisely its old set of colors. When two tokens stop
before $w$, their last leaves both point to $w$ and hence miss the color
of $sw$. This gives the two-blank configuration used in separable
fans~\cite{ABBCSZ26R}. At most one token enters any tree vertex. The
cycle arcs used by pairs have disjoint vertices, and tokens waiting
before their entries occupy off-cycle leaves. In the unique-entry case
both $y$ and $z$ point to $w$. A self-loop would put the mark at $w$
on its own spoke $sw$, so the cycle has length at least two and
$y,z,w$ are distinct. These facts give all the asserted disjointness, including the long prefix of
each nested fan. Only a cycle with at least three unpaired arrivals loses
a token. At least two thirds of all tokens are used, at most two per choice.

We can compute the token paths without passing tokens sequentially.
In each in-tree the parity of the number of blank starts in a child
subtree says whether that child sends a token. Reachability after cutting
the cycle edges identifies the subtrees, and local prefix sums pair the
odd children. The unique continuing child, if any, gives a pointer from
a parent to the surviving start. Doubling pointers recovers the paths.
On the cycles we rank entries from the least identifier and use their
ranks to recover the prescribed arcs. Boolean reachability on at most
$q-1$ vertices and sorting lists of this size take $O(L^2)$ depth with
$O(q^3)$ processors per center. Even with all centers active,
$nq^3=O(n^3q)$.
\end{proof}

The next lemma concerns a single choice in the coloring in which it
rotates. For independent centers with disjoint leaf sets, we may first
fix and apply every other rotation and exit fill. For choices sharing
one center, the color reservations in Lemma~\ref{lem:restore} give the
required missing colors.

\begin{lemma}[Usable terminals]\label{lem:terminal}
After a two-blank rotation, if $s$ misses $a$ and both terminals miss
$b$, at least one terminal is usable. For a nested fan with terminal
mark $b$ and missing center color $a$, the short prefix gives a usable
terminal if its leaf $y$ lies outside the pre-rotation $(a,b)$-component
of $s$. Otherwise the long prefix gives a usable terminal.
\end{lemma}
\begin{proof}
For two blanks, the rotation preserves the colors present at $s$ and
vacates the two terminal spokes. The center still misses $a$ and both
leaves still miss $b$. The component of the
center is a path with the center as one endpoint, or an isolate. It cannot
contain both other endpoints.

For a nested fan, the short rotation moves neither $a$ nor $b$. If its
terminal $y$ is outside the center component, we are done. Otherwise the
component is a path from $s$ to $y$, starting with the $b$-spoke $sw$.
The long rotation removes $sw$ from the two-color subgraph and attaches
$s$ at the other end by the $b$-spoke $sy$. These are its only changes
to that subgraph, so the component keeps the same vertices. Before this
change $z$ was distinct from both endpoints and missed $b$, whereas
every internal vertex of the path had an incident $b$-edge. Thus $z$
stays outside the path and still misses $b$
(Figure~\ref{fig:prefix}).
\end{proof}

\section{Choosing simultaneous rotations}\label{sec:choices}
We now implement one phase at several centers. The input choices have
disjoint leaves and distinct ordered pairs at each center. We first
select fans with disjoint leaves across centers, then choose their
rotation lengths. In the resulting coloring we fill as many designated
blanks as their assigned pairs permit.

\subsection{Algorithm}
\begin{algorithm}[H]
\caption{One phase at several centers}\label{alg:joint}
\io{The current coloring, the marks, and the choices retained by Algorithm~\ref{alg:restore}, with their ordered pairs.}
   {A proper partial coloring on the same palette with fewer blanks whenever a blank remains.}
\noindent Set $H=1 + 2\tau + B'$, let $P$ be the least prime at least
$4\max\{n,H\}$, and set $r=\floor{P/(2H)}/P$.
\begin{steps}
\st{For each oriented palette matching below, activate its ordered pairs. Keep the matching maximizing $r|A| - 2r^2\,e(A)$, where $A$ is the set of active choices and $e(A)$ counts unordered pairs of choices in $A$ that share a leaf.}
\st{For the winning matching, try each affine thinning map. Keep the largest set of selected choices that meet no other selected choice.}
\st{Rotate and fill its exits, and rotate both paths at its two-blank choices.}
\st{Try the prefix assignments below on separate copies. Rotate each nested fan by its assigned length. Count the nondirect choices with a usable terminal in that coloring. Commit the best assignment.}
\st{Designate one usable terminal per successful nondirect choice. For each matched pair $\{a,b\}$, form $H_{a,b}$ from its designated blanks and all edges colored $a$ or $b$.}
\st{Leave the least designated blank on each odd cycle unfilled. Color the remaining paths and even cycles of each $H_{a,b}$ alternately with $a$ and $b$.}
\st{Commit the resulting component exchanges, fill the retained blanks with their computed colors, and return the coloring.}
\end{steps}
\end{algorithm}

\paragraph{Selecting disjoint fans.}
Let $p$ be the least prime at least $\max\{5,q\}$. We use the palette
labels as elements of $\F_p$. For $c\in\F_p$ and $z\in\F_p^\times$,
we match distinct labels $x,y$ satisfying
\begin{equation}\label{eq:pairs}
                         (x-c)(y-c)=z .
\end{equation}
A common bit orients all pairs in increasing or decreasing label order.
Each label has at most one partner. Each center uses a single first color
$a_s$ and has at most one choice for each second color, so at most one
choice is active there. Two active choices conflict when they share a
leaf. The centers belong to an independent set, hence two choices with
no shared leaf have disjoint vertex sets.

For thinning we set $H=1+2\tau+B'$ and take a prime $P=\Theta(\max\{n,H\})$
with $P\ge4\max\{n,H\}$. We select a center when its affine residue
is below $\floor{P/(2H)}$. Thus the rate in both scoring steps is the
same rational number $r=\floor{P/(2H)}/P$. It satisfies
$r=\Theta(1/H)$ and $rH\le1/2$.

\paragraph{Trying prefix lengths.}
Bit zero chooses the short prefix, and bit one the long prefix. With
no nested fans we use the empty assignment. Otherwise let $N$ be their
number and set $K=\ceil{\log_2 n}+C$ for a sufficiently large fixed
integer $C$. We use the algebraic small-bias
construction of Alon et al.~\cite{AGHP92}. We index the fans by distinct
elements $z_i$ of $\F_{2^u}$ for $u=\ceil{\log_2(N+1)}$. For each fan
let $v_i$ be the concatenated binary coordinates of
$(1,z_i,\ldots,z_i^{K-1})$, of length $d_0=Ku$. Let $M_0$ be the least
power of two at least $C(d_0+1)2^K$. A trial uses $x,y\in\F_{M_0}$
and assigns fan $i$ the bit
\begin{equation}\label{eq:bias}
 w_j=\langle y,x^j\rangle_2\quad(0\le j<d_0),\qquad
 b_i=\langle v_i,w\rangle_2 .
\end{equation}
Angle brackets denote the dot product of binary coordinate vectors.
We score assignments by computing the actual components after rotation.
The query bound $K$ is used only to prove that some trial scores well.

\subsection{Analysis}
\paragraph{How many fans survive selection.}
For a specified ordered pair $(x,y)$, each $c\notin\{x,y\}$ forces a
nonzero $z$ and one orientation. The pair is therefore active with
probability $(p-2)/(2p(p-1))=\Theta(1/q)$. For two specified
disjoint ordered pairs, equating the two products in~\eqref{eq:pairs}
gives a linear equation in $c$. If its linear coefficient vanishes,
a solution would give equal sums and products, hence the same unordered
pair. Otherwise there is at most one $c$, and at most one orientation
works. Conditional on the first pair being active, the second is thus
active with probability at most $1/(p-2)$.

Condition on the activation of one choice at $s$. It has at most $q-1$
leaves and each leaf belongs to at most $\tau$ choices, one per adjacent
center. The expected number of conflicts with disjoint color pairs is
at most $(q-1)\tau/(p-2)\le2\tau$. Distinct ordered pairs sharing a label
cannot be selected together. Conflicts with the identical pair come
from centers in $W_s$ whose mark is $a_s$, of which there are at most $B'$.
Consequently $\E[2e(A)]\le H\,\E|A|$.

There are $\Omega(d/B)$ input choices, so $\E|A|=\Omega(d/(qB))$.
For each fixed $A$, pairwise independent thinning selects at least
$r|A|-2r^2e(A)$ isolated choices in expectation. Averaging
the score over matchings gives at least $r\,\E|A|/2$ because $rH\le1/2$.
Choosing the best matching by this score and then the best thinning
map for that matching yields a vertex-disjoint batch of size
\begin{equation}\label{eq:batch}
                    \Omega\!\left(\frac{d}{qB(1+\tau+B')}\right).
\end{equation}

\begin{figure}[tbp]
\centering
\begin{tikzpicture}[
  x=1cm,y=1cm,font=\small,>=Stealth,
  v/.style={circle,draw=black!75,fill=white,inner sep=0pt,
            minimum size=4.4pt,line width=.65pt},
  ctr/.style={circle,draw=ecD!90!black,fill=ecD!18,inner sep=0pt,
            minimum size=5.4pt,line width=.85pt},
  eA/.style={draw=ecA,line width=1.55pt,line cap=round},
  eB/.style={draw=ecB,line width=1.55pt,line cap=round},
  eAlt/.style={draw=ecB!60,line width=1.25pt,densely dotted,line cap=round},
  fanbox/.style={fill=ecD!8,draw=ecD!45,rounded corners=4pt,line width=.6pt},
  lab/.style={font=\scriptsize,inner sep=1.2pt}
]
% Panel A (left): Center lookup vs. leaf lookup.
\begin{scope}[xshift=0cm]
  \node[font=\small] at (1.60,2.15) {Fan $t$ of a different pair};

  % Top sub-box: lookup at center t (both bit values continue).
  \begin{scope}[yshift=0.95cm]
    \coordinate (in1) at (0,0);
    \coordinate (t1) at (1.10,0);
    \coordinate (u0) at (2.60,0.48);
    \coordinate (u1) at (2.60,-0.48);
    \draw[eA,->] (in1)--node[lab,above=1pt,text=ecA] {$a_s$} ($(t1)+(-.12,0)$);
    \draw[eB,->] (t1)--node[lab,pos=.58,above,sloped,text=ecB] {$\beta_t{=}0$} (u0);
    \draw[eAlt,->] (t1)--node[lab,pos=.58,below,sloped,text=ecB!85!black] {$\beta_t{=}1$} (u1);
    \node[v] at (in1) {};
    \node[ctr] at (t1) {};
    \node[v] at (u0) {};
    \node[v] at (u1) {};
    \node[lab,below=4pt,text=ecD!85!black] at (t1) {$t$};
    \node[lab,right=2pt] at (u0) {$u_0$};
    \node[lab,right=2pt] at (u1) {$u_1$};
    \node[lab,text=black!65] at (1.50,-0.88) {both values select a $b$-spoke};
  \end{scope}

  % Bottom sub-box: lookup at leaf u0 (one bit enters t, other stops).
  \begin{scope}[yshift=-0.95cm]
    \coordinate (in2) at (0,0);
    \coordinate (u0b) at (1.10,0);
    \coordinate (t2) at (2.60,0.44);
    \coordinate (stp) at (2.30,-0.44);
    \draw[eA,->] (in2)--node[lab,above=1pt,text=ecA] {$a_s$} ($(u0b)+(-.12,0)$);
    \draw[eB,->] (u0b)--node[lab,pos=.58,above,sloped,text=ecB] {$\beta_t{=}0$} (t2);
    \draw[->,black!45,densely dotted,line width=.9pt]
      (u0b)--node[lab,pos=.56,below,sloped,text=black!60] {$\beta_t{=}1$} (stp);
    % Stop cross at stp.
    \draw[ecA!85!black,line width=.85pt]
      ($(stp)+(-.09,-.09)$)--($(stp)+(.09,.09)$)
      ($(stp)+(-.09,.09)$)--($(stp)+(.09,-.09)$);
    \node[v] at (in2) {};
    \node[v] at (u0b) {};
    \node[ctr] at (t2) {};
    \node[lab,below=4pt] at (u0b) {$u_0$};
    \node[lab,right=3pt,text=ecD!85!black] at (t2) {$t$};
    \node[lab,right=4pt,text=ecA!85!black] at (stp) {stop};
    \node[lab,text=ecC!70!black] at (1.50,-0.88) {leaf: passes with prob.\ $1/2$};
  \end{scope}
\end{scope}

% Subtle vertical separator.
\draw[black!18,line width=.6pt] (3.85,-1.95)--(3.85,2.25);

% Panel B (right): Realized path traced from both ends.
\begin{scope}[xshift=4.65cm]
  \node[font=\small] at (4.65,2.15) {Charging bit-dependent spokes along the realized path};

  % Coordinates of the 8-vertex alternating path.
  \coordinate (s)  at (0.00,0);
  \coordinate (v1) at (1.30,0);
  \coordinate (u1) at (2.60,0);
  \coordinate (t1) at (3.95,0);
  \coordinate (v2) at (5.30,0);
  \coordinate (t2) at (6.65,0);
  \coordinate (u2) at (8.00,0);
  \coordinate (e)  at (9.30,0);

  % Soft rounded boxes around the two interior variable fans.
  \draw[fanbox] (2.25,-0.55) rectangle (4.30,0.52);
  \draw[fanbox] (6.30,-0.55) rectangle (8.35,0.52);
  \node[lab,text=ecD!85!black] at (3.28,-0.74) {fan $t_1$};
  \node[lab,text=ecD!85!black] at (7.33,0.72) {fan $t_2$};

  % Alternating edges along realized path: b, a_s, b, a_s, b, a_s, b.
  \draw[eB] (s)--node[lab,above=1pt,text=ecB] {$b$} (v1);
  \draw[eA] (v1)--node[lab,above=1pt,text=ecA] {$a_s$} (u1);
  \draw[eB,line width=2.25pt] (u1)--node[lab,above=1pt,text=ecB] {$b$} (t1);
  \draw[eA] (t1)--node[lab,above=1pt,text=ecA] {$a_s$} (v2);
  \draw[eB] (v2)--node[lab,above=1pt,text=ecB] {$b$} (t2);
  \draw[eA,line width=2.25pt] (t2)--node[lab,above=1pt,text=ecA] {$a_s$} (u2);
  \draw[eB] (u2)--node[lab,above=1pt,text=ecB] {$b$} (e);

  % Directional charging arrows cleanly outside the fan boxes.
  \draw[->,ecC!80!black,line width=1.1pt] (2.72,0.78)--(3.83,0.78);
  \node[lab,above=2pt,text=ecC!75!black] at (3.28,0.80) {forward $1/2$ test};

  \draw[->,ecC!80!black,line width=1.1pt] (7.88,-0.78)--(6.77,-0.78);
  \node[lab,below=2pt,text=ecC!75!black] at (7.33,-0.80) {reverse $1/2$ test};

  % Vertices.
  \foreach \p in {s,v1,u1,v2,u2,e} {\node[v] at (\p) {};}
  \node[ctr] at (t1) {};
  \node[ctr] at (t2) {};

  \node[lab,below=3pt] at (s) {$s$};
  \node[lab,below=3pt] at (u1) {$u_1$};
  \node[lab,below=3pt,text=ecD!85!black] at (t1) {$t_1$};
  \node[lab,below=3pt,text=ecD!85!black] at (t2) {$t_2$};
  \node[lab,below=3pt] at (u2) {$u_2$};
  \node[lab,below=3pt] at (e) {$e$};

  % Global forward / reverse sweep indicators.
  \draw[->,black!60,line width=.8pt] (0.15,1.58)--(9.15,1.58);
  \node[lab,fill=white,inner sep=2.5pt,text=black!75] at (4.65,1.58)
    {forward trace from $s$};

  \draw[->,black!60,line width=.8pt] (9.15,-1.58)--(0.15,-1.58);
  \node[lab,fill=white,inner sep=2.5pt,text=black!75] at (4.65,-1.58)
    {reverse trace from $e$ (at most $2n$ possible starts)};
\end{scope}
\end{tikzpicture}
\caption{Bidirectional charging in Lemma~\ref{lem:expose}. The interior
fan centers shown are scheduled for pairs disjoint from $\{a_s,b\}$.
Left: querying a fresh fan bit $\beta_t$ at center $t$ chooses between two spokes,
whereas querying it at leaf $u_0$ enters $t$ with probability $1/2$ and
terminates the walk otherwise. Right: along the realized $(a_s,b)$-path
from $s$ to $e$, every interior bit-dependent spoke (bold) is traversed
toward its center either forward ($u_1\to t_1$) or in reverse
($u_2\to t_2$), triggering a fresh $1/2$ leaf test in that direction.}
\label{fig:both-ends}
\end{figure}
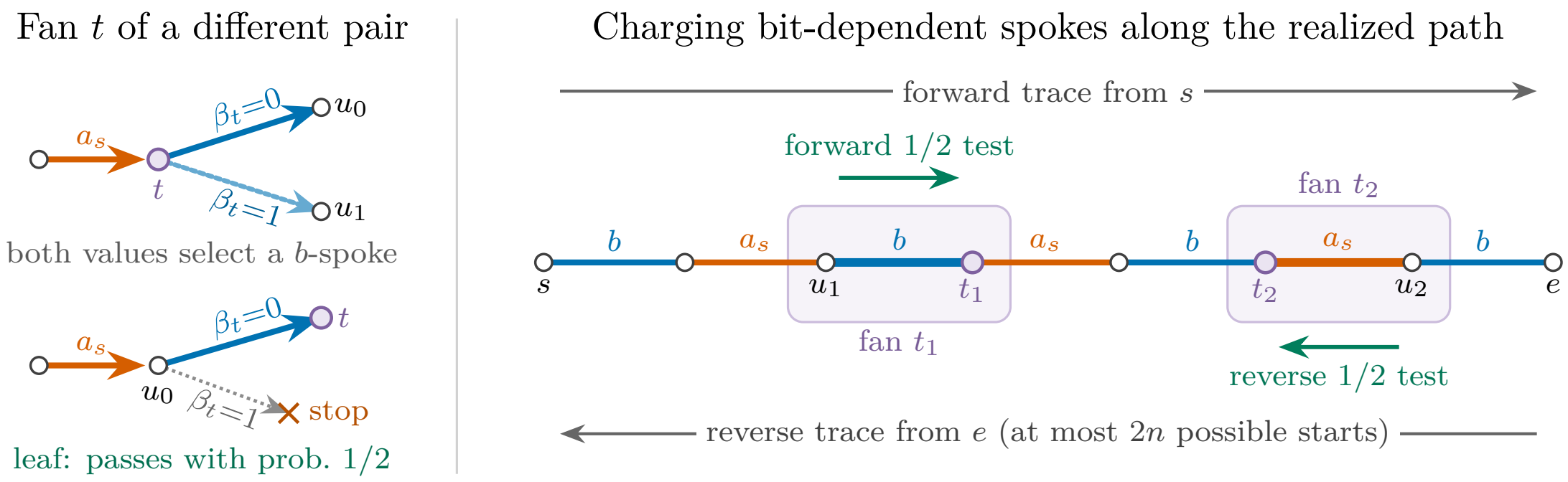

\paragraph{What a path reveals.}
The $(a_s,b)$-path can cross the center of a nested fan scheduled for a
different pair. For example, its long prefix can move $b$ to the preceding
spoke while leaving an $a_s$-spoke in place. A lookup at that interior
center can take either of two $b$-spokes. The proof
charges a bit to the direction in which one of its changed spokes is
traversed toward its center, where reaching the center required a
successful leaf test (Figure~\ref{fig:both-ends}).

\begin{lemma}[Reading an alternating path from both ends]\label{lem:expose}
Fix a nested fan at $s$ and its own bit. Give the other nested fans
independent fair bits. Starting at $s$ with color $b$, we follow its
$(a_s,b)$-path, inspecting a bit only when the next edge depends on it.
For integers $f,g\ge1$, the probability of inspecting more than $f+g+4$
distinct other bits is at most $2^{-f}+2n\,2^{-g}$.
\end{lemma}
\begin{proof}
The colors present at a variable fan center are the same for both bit
values. In particular, the center of a nested fan scheduled for $(a_s,b)$
still misses $a_s$ and has degree at most one in the traced subgraph.
A lookup at a center whose answer varies chooses between two spokes.
At a leaf only its spoke to that center changes. If a lookup at the leaf
depends on the bit, one value takes that spoke and the other says that
the requested color is absent. Since the local vertex sets are disjoint,
we can determine whether the two possible answers agree before reading
an unknown bit.

For any fixed starting vertex and first requested color, the next such
lookup at a leaf with an unread bit succeeds with probability $1/2$
conditional on the entire history. Failure stops the walk. By successive
conditioning, the probability of passing $h$ fresh leaf tests is at most
$2^{-h}$. The bound holds if we also stop on the first repeated vertex.

The realized path from $s$ is simple since $s$ misses $a_s$. Except for
a possible terminating lookup that finds no edge, a first read at a leaf
reaches its center, and a first read at a center takes a spoke that changes
with the bit. Thus each interior center so read has a bit-dependent spoke
on the path. In the direction traversing this spoke toward its center,
the lookup at its leaf is a fresh successful test. If the bit had been
read earlier in this traversal, the walk would already have visited the
center or would have stopped. Distinct centers give distinct tests.
The two endpoints and a possible terminating lookup account for at most
three further bits. With fewer than $f$ forward and $g$ reverse successes
there are at most $(f-1)+(g-1)+3=f+g+1$ reads, within the stated bound.
The forward instruction is fixed. For each fixed vertex $v$ and first
color $c\in\{a_s,b\}$, let $E_{v,c}$ be the event that the walk from
$(v,c)$ passes $g$ fresh leaf tests. Each has probability at most $2^{-g}$.
If the realized reverse walk passes $g$ tests, its starting vertex $e$
and first color $c$ place the assignment in $E_{e,c}$. Taking the union over all $2|V|\le 2n$ instructions
therefore bounds this probability by $2n\,2^{-g}$.
\end{proof}

\paragraph{Why the finite list suffices.}
For a given nested fan, its leaf still misses $b$ and the center still
misses $a_s$. The leaf is unusable precisely when it is the far endpoint
of the center's alternating path. Under independent bits the probability
of a usable leaf is at least $1/2$ by Lemma~\ref{lem:terminal}, even
after fixing all the other rotations. For the analysis we test a fan by
first reading its own bit and then tracing the path. We truncate a test
as a rejection if another query would exceed $K$. Taking $f$ constant
and $g=\ceil{\log_2 n}+f$ in Lemma~\ref{lem:expose} gives a tail of
at most $3\cdot2^{-f}$. We choose $f$ large enough to make this at most
$1/16$, then $C$ large enough to include these queries and the fan's own bit.

The truncated test has at most $2^K=O(n)$ accepting leaves as a binary
decision tree. Each leaf specifies at most $K$ distinct literals. We
show that the probability of each such conjunction under~\eqref{eq:bias}
differs from its independent-bit probability by at most $(d_0-1)/M_0$.
For $k\le K$ distinct fan indices $z_i$, the first $k$ powers form a nonsingular
Vandermonde matrix. Hence every nonempty sum of at most $K$ distinct
$v_i$ is nonzero. Every nonzero parity of $w$ has bias at most $(d_0-1)/M_0$ in
absolute value. Indeed, averaging over $y$ gives zero unless $x$ is a root
of the corresponding nonzero binary-coefficient polynomial of degree at
most $d_0-1$. Thus the same bias bound holds for every relevant nonempty
parity of the $b_i$. The indicator of a conjunction expands as a product
of factors $(1\pm(-1)^{b_i})/2$ whose Fourier coefficients have total
absolute value one. This proves the claim. Summing over the accepting
leaves costs at most $2^K(d_0-1)/M_0\le 1/C$. Acceptance after truncation
always certifies a usable terminal, so each nested fan succeeds on at
least a third of the explicit assignments.

\paragraph{Completing the fixed pairs.}
Fix a proper partial coloring and assign each designated blank to one of
several pairwise disjoint color pairs. For each pair $\{a,b\}$, we add its
blanks to its two-color subgraph to obtain $H_{a,b}$. We may switch whole
original components and fill designated blanks with either color of their
pair. If each end misses exactly one of $a,b$ and these colors differ, the blank is
individually fillable exactly when the two component switch bits differ.

\begin{lemma}[Fixed-pair completion]\label{lem:fixed-pair}
All designated blanks can be filled if and only if every $H_{a,b}$ has maximum
degree at most two and no odd cycle. Under the degree condition a
maximum-cardinality completion omits one blank per odd cycle and can be
found in deterministic logspace.

If the blanks form a matching and each joins distinct original components
of its pair, with one color missing at one end and the other color missing
at the other end, at least two thirds of the designated blanks can be filled.
\end{lemma}
\begin{proof}
A completion of all blanks gives a proper two-edge-coloring of each $H_{a,b}$. Conversely,
alternating the colors on paths and even cycles either preserves or
exchanges the coloring on each original component. Distinct pairs use
disjoint colors, so their completions are compatible at shared vertices. Under the
degree condition each odd cycle requires an omission and contains a blank,
since its old edges were properly colored. Omitting any such blank leaves a path.

Under the matching and missing-color hypotheses the degrees are at most two.
On a cycle each end of a blank has one old incident edge. The opposite
missing colors force these two edges to have opposite colors.
Skipping the blanks leaves a cyclically alternating sequence of old edges,
of even length. An odd cycle therefore contains an odd number of blanks.
A single blank would have both ends on the remaining old path. Thus every
odd cycle contains at least three blanks.

Rescanning the input gives degrees and neighbors. For each output edge we
walk its component, recording its least designated blank if it is an odd
cycle. We start each remaining path at its smaller endpoint and each
even cycle at its least vertex toward
its smaller neighbor. Alternating from the smaller color gives the edge's
color using only a constant number of identifiers and counters.
\end{proof}
For nonnegative weights and degrees at most two, omitting a lightest blank
per odd cycle is optimal. With at least three blanks per odd cycle we
retain at least two thirds by weight.

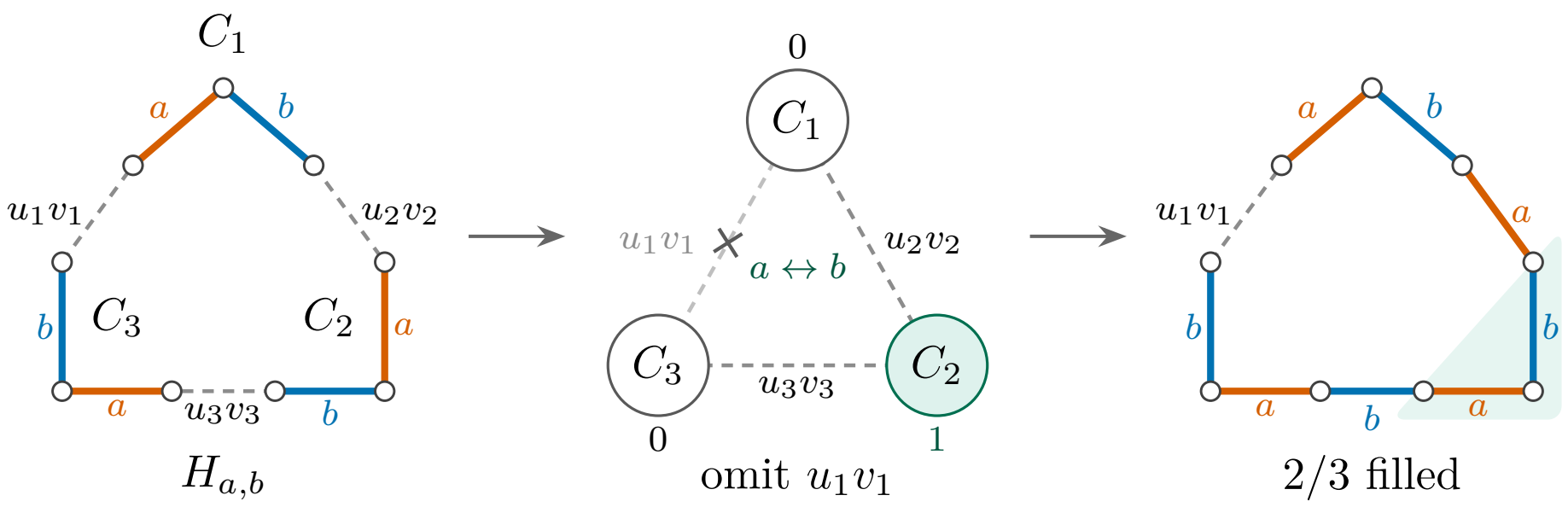
\begin{figure}[tbp]
\centering
\begin{tikzpicture}[
  x=1cm,y=1cm,font=\small,
  vertex/.style={circle,draw=black!75,fill=white,line width=.55pt,
    inner sep=0pt,minimum size=4.2pt},
  aedge/.style={draw=ecA,line width=1.55pt,line cap=round},
  bedge/.style={draw=ecB,line width=1.55pt,line cap=round},
  blank/.style={draw=black!45,line width=.85pt,dash pattern=on 2.5pt off 2pt},
  arrow/.style={-{Stealth[length=2.2mm]},draw=black!60,line width=.75pt},
  component/.style={circle,draw=black!65,fill=white,line width=.6pt,
    inner sep=2.2pt,minimum size=7.8mm},
  lab/.style={font=\scriptsize,inner sep=1.2pt}
]
% Left panel: augmented subgraph H_{a,b} before completion.
\begin{scope}
  \coordinate (v1) at (-.70,.90);
  \coordinate (w1) at (0,1.50);
  \coordinate (u2) at (.70,.90);
  \coordinate (v2) at (1.25,.15);
  \coordinate (w2) at (1.25,-.85);
  \coordinate (u3) at (.40,-.85);
  \coordinate (v3) at (-.40,-.85);
  \coordinate (w3) at (-1.25,-.85);
  \coordinate (u1) at (-1.25,.15);
  \draw[aedge] (v1)--node[lab,above left,text=ecA] {$a$} (w1);
  \draw[bedge] (w1)--node[lab,above right,text=ecB] {$b$} (u2);
  \draw[aedge] (v2)--node[lab,right,text=ecA] {$a$} (w2);
  \draw[bedge] (w2)--node[lab,below,text=ecB] {$b$} (u3);
  \draw[aedge] (v3)--node[lab,below,text=ecA] {$a$} (w3);
  \draw[bedge] (w3)--node[lab,left,text=ecB] {$b$} (u1);
  \draw[blank] (u2)--node[lab,right=1pt] {$u_2v_2$} (v2);
  \draw[blank] (u3)--node[lab,below=1pt] {$u_3v_3$} (v3);
  \draw[blank] (u1)--node[lab,left=1pt] {$u_1v_1$} (v1);
  \foreach \p in {v1,w1,u2,v2,w2,u3,v3,w3,u1}
    \node[vertex] at (\p) {};
  \node at (0,1.92) {$C_1$};
  \node at (.82,-.28) {$C_2$};
  \node at (-.82,-.28) {$C_3$};
  \node at (0,-1.52) {$H_{a,b}$};
\end{scope}

\draw[arrow] (1.90,.35)--(2.65,.35);

% Middle panel: component switch view, omitting least blank u_1v_1.
\begin{scope}[xshift=4.45cm]
  \coordinate (c1) at (0,1.25);
  \coordinate (c2) at (1.08,-.65);
  \coordinate (c3) at (-1.08,-.65);
  \draw[blank] (c1)--node[lab,right=2pt] {$u_2v_2$} (c2);
  \draw[blank] (c2)--node[lab,below=1pt] {$u_3v_3$} (c3);
  \draw[blank,draw=black!25] (c3)--node[lab,left=5pt,text=black!45] {$u_1v_1$} (c1);
  \coordinate (cut) at ($(c3)!.5!(c1)$);
  \draw[black!65,line width=.8pt]
    ($(cut)+(-.11,-.07)$)--($(cut)+(.11,.07)$)
    ($(cut)+(-.07,.11)$)--($(cut)+(.07,-.11)$);
  \node[component] at (c1) {$C_1$};
  \node[component,fill=ecC!13,draw=ecC!70!black] at (c2) {$C_2$};
  \node[component] at (c3) {$C_3$};
  \node[lab,above=1pt] at ($(c1)+(0,.40)$) {$0$};
  \node[lab,below=2pt] at ($(c3)+(0,-.36)$) {$0$};
  \node[lab,below=2pt,text=ecC!55!black] at ($(c2)+(0,-.36)$) {$1$};
  \node[lab,text=ecC!55!black] at (0,.12) {$a\leftrightarrow b$};
  \node at (0,-1.52) {omit $u_1v_1$};
\end{scope}

\draw[arrow] (6.25,.35)--(7.00,.35);

% Right panel: completed alternating path after swapping C_2.
\begin{scope}[xshift=8.90cm]
  \coordinate (v1f) at (-.70,.90);
  \coordinate (w1f) at (0,1.50);
  \coordinate (u2f) at (.70,.90);
  \coordinate (v2f) at (1.25,.15);
  \coordinate (w2f) at (1.25,-.85);
  \coordinate (u3f) at (.40,-.85);
  \coordinate (v3f) at (-.40,-.85);
  \coordinate (w3f) at (-1.25,-.85);
  \coordinate (u1f) at (-1.25,.15);
  \fill[ecC!10,rounded corners=3pt]
    (1.47,.40)--(1.47,-1.07)--(.15,-1.07)--cycle;
  \draw[aedge] (v1f)--node[lab,above left,text=ecA] {$a$} (w1f);
  \draw[bedge] (w1f)--node[lab,above right,text=ecB] {$b$} (u2f);
  \draw[bedge] (v2f)--node[lab,right,text=ecB] {$b$} (w2f);
  \draw[aedge] (w2f)--node[lab,below,text=ecA] {$a$} (u3f);
  \draw[aedge] (v3f)--node[lab,below,text=ecA] {$a$} (w3f);
  \draw[bedge] (w3f)--node[lab,left,text=ecB] {$b$} (u1f);
  \draw[aedge] (u2f)--node[lab,right=1pt,text=ecA] {$a$} (v2f);
  \draw[bedge] (u3f)--node[lab,below=1pt,text=ecB] {$b$} (v3f);
  \draw[blank] (u1f)--node[lab,left=1pt] {$u_1v_1$} (v1f);
  \foreach \p in {v1f,w1f,u2f,v2f,w2f,u3f,v3f,w3f,u1f}
    \node[vertex] at (\p) {};
  \node at (0,-1.52) {$2/3$ filled};
\end{scope}
\end{tikzpicture}
\caption{Fixed-pair completion on an odd cycle of $H_{a,b}$
(Lemma~\ref{lem:fixed-pair}). Three original two-edge $(a,b)$-paths
$C_1,C_2,C_3$ are joined by designated blanks $u_1v_1,u_2v_2,u_3v_3$ into a
nine-cycle. Omitting the least blank $u_1v_1$ and exchanging $a\leftrightarrow b$
on $C_2$ fills $u_2v_2$ with $a$ and $u_3v_3$ with $b$, attaining the sharp
$2/3$ fraction.}
\label{fig:fixed-pair}
\end{figure}

The fraction is sharp for fixed pairs. For $i$ modulo three,
leave $u_i v_i$ blank and give $v_iw_i$ and $w_i u_{i+1}$ colors $a$ and
$b$, respectively. Each blank joins distinct original paths, and the
augmented nine-cycle forces one omission (Figure~\ref{fig:fixed-pair}).

\begin{lemma}[Progress at several centers]\label{lem:joint}
On the choices produced in a phase of Algorithm~\ref{alg:restore} with
$d>0$ blanks, Algorithm~\ref{alg:joint} fills
$\Omega(d/(qB(1+\tau+B')))$ blanks. It preserves properness and the
centers of all remaining blanks. Its depth is $O(L^{10})$ and it uses
$O(n^3q)$ processors.
\end{lemma}
\begin{proof}
Every exit in the selected batch is filled and stays colored under the
subsequent component exchanges. Every two-blank choice has
a usable terminal by Lemma~\ref{lem:terminal}, and each nested fan
contributes one with probability at least $1/3$ on the prefix list.
Thus the best prefix assignment designates at least a third of the
nondirect choices. Their blanks form a matching because the selected
fans are vertex-disjoint. The center misses $a_s$, its terminal misses
$b$, and usability puts them in distinct components in the coloring
after rotation. Lemma~\ref{lem:fixed-pair} fills at least two thirds of
these blanks. Including the exits, at least $2/9$ of the selected batch
is filled. Equation~\eqref{eq:batch} now gives the claimed decrease. Only
rotations move blanks, and each moves along spokes of its own center.

The processor and depth bounds come from the following steps.
\begin{itemize}[leftmargin=*,itemsep=3pt]
\item \textbf{Selecting the batch.} At each leaf at most $\tau$ choices
meet. We list their pairs and remove duplicates by sorting, giving
$O(n\tau^2)$ potential conflicts. For each conflict and $c$, the first
choice forces $z$ and the orientation in~\eqref{eq:pairs}, if it occurs.
Testing the second choice and summing records by seed counts $e(A)$.
Single-choice records similarly give $|A|$. We compare matching scores
after multiplying by their common denominator $P^2$. These tables fit
$O(n\tau^2q+nq^2)=O(n^3q)$ processors. For the winning matching there
are $O(n^2)$ thinning seeds because $H=O(n)$. For each seed we sort
$O(nq)$ leaf incidences of marked choices to flag conflicts. A rational
score can be compared after multiplication by the common denominator $P^2$.
\item \textbf{Trying prefix lengths.} Since $N\le n$, $d_0=O(L^2)$ and
$M_0=O(nL^2)$. There are $O(n^2L^4)$ trials. We run $O(n^2)$ at a time
and keep the best seed across batches. Each trial uses $O(nq)$ processors.
Each fan can form its own bit using $O(L^2)$ field powers and dot products
on $O(L)$-bit elements, taking $O(L^6)$ time per fan computation. Each edge reads its new
color from its local rotation, if any. Properness gives distinct
vertex--color incidence slots, so we can write this table directly.
For each matched pair, pointer doubling finds component identifiers,
including for isolates, in $O(L)$ depth. Scoring the usable terminals by
these identifiers takes a reduction. The $O(L^4)$ batches account for
the asserted $O(L^{10})$ depth. For each field size $M\in\{2^u,M_0\}$, we
enumerate monic polynomials of degree $\log_2 M$ and test division by
all monic factors up to half that degree. The least irreducible polynomial
fixes the representation. This takes $O(M^{3/2})$ trials and $O(L^2)$
depth. Even for the larger field the $O(n^{3/2}L^3)$ trials fit within
the $O(n^3q)$ processor budget. Prime finding, sorting, and sums
over records take $O(L^3)$ depth within the same processor budget.
\item \textbf{Completing the pairs.} The incidence table supplies the
colored neighbors at each vertex for each pair. The designated matching
adds at most one neighbor. In each $H_{a,b}$, pointer doubling finds path
endpoints and component minima, including the least designated blank.
Cutting a cyclic list at its least vertex in the canonical direction
lets us rank its edges and test the parity of its length. After the
odd-cycle omissions, a second ranking gives the alternating colors.
Each edge writes its own color. All pairs together use $O(nq)$
processors and $O(L)$ depth on CREW.
\end{itemize}
\end{proof}

\section{Balancing integer rows on a CREW PRAM}\label{sec:balance}
The last routine supplies~\eqref{eq:sign}. In one application a row counts vertices in a
neighborhood. In the other it counts marks at the ends of alternating paths. We give the bounded-word construction to make
explicit the cost of satisfying all of these constraints at once.

\subsection{Algorithm}
The input consists of the $M$ nonzero entries of an $R$-by-$s$ integer
matrix with entries of absolute value at most two. We discard zero
rows. If none remain, any signs suffice. Thus we may take $R,s\ge1$. We use the
absorbing-walk method of Ghaffari, Grunau, and Rozho\v{n}~\cite{GGRwalk}.
Each column starts at position zero and takes steps of size one until
it reaches $-k$ or $k$. For a sufficiently large fixed integer $c$ we use
\[
 k=cL\ceil{\log_2(2R)},\qquad
 T=2ck^2\ceil{\log_2(2s)},\qquad J=2RT.
\]
Let $H_t(x)$ be the probability that a fair walk at $x$ at time $t$
is unabsorbed at time $T$. We precompute this table by backward averaging
with zero boundary values and $H_T(x)=\mathbf{1}_{|x|<k}$.

For a nonzero row $h$ let $A=\sum_j|h_j|$, $V=\sum_jh_j^2$ and
$\lambda=1/(2A)$. The row has two scores $P_+,P_-$, both initially one.
At each of the $T$ steps we try all affine binary forms on the column
identifiers, using their bits as increments in $\{-1,1\}$ on unabsorbed
columns. Absorbed columns take increment zero. For a trial let $x'_j$
be the proposed column positions and $Y$ the sum of the increments
weighted by $h$. We obtain the proposed row scores $P'_\pm$ by rounding
\begin{equation}\label{eq:step}
               P_\pm\,\frac{1\pm\lambda Y+\lambda^2Y^2}{1+\lambda^2V}
\end{equation}
up to the next multiple of $1/J$, retaining an exact multiple.
We commit the trial minimizing
\[
               \sum_h(P'_++P'_-)+4R\sum_j H_{t+1}(x'_j).
\]
At time $T$ the algorithm returns $\sigma_j=x_j/k$.

\subsection{Analysis}
The two parts of the score have different jobs. A large signed row sum
forces one of its row scores up. A column that survives to the final
step incurs a penalty exceeding the entire allowed score. Pairwise
independence at one step suffices to bound the row scores, and a single
walk's recurrence controls that column's penalty.

\begin{lemma}[Simultaneous row balancing]\label{lem:sign}
For $R,s,M$ polynomial in $n$, the algorithm returns signs in $\{-1,1\}$
satisfying~\eqref{eq:sign} for every row. Its deterministic CREW depth is
$O(L^{12})$ with $O(s(M+R+s))$ processors on $O(L)$-bit words when $s\ge1$.
\end{lemma}
\begin{proof}
The expected absorption time of a fair walk at $x$ is $k^2-x^2$, by the
two-neighbor recurrence with zero boundary values. From any interior
starting point a block of $2k^2$ steps has survival probability at most
$1/2$ by Markov's inequality. Applying this to successive blocks gives
$H_0(0)\le1/(16s)$ for our choice of $T$.

Condition on the entire state at a step. Over a uniform trial, the
unabsorbed increments are fair and pairwise independent. Thus
$\E[Y]=0$, $\E[Y^2]\le V$, and each unrounded score in~\eqref{eq:step}
has expectation at most its old value. Rounding adds at most $2R/J\le1/T$
to the expected sum. By the defining recurrence of $H_t$, the expected
walk penalty equals its current value. The selected trial can cost no
more than this average. The initial score is at most $2R+R/4$,
and it increases by at most one over all $T$ steps. It stays strictly
below $4R$, so at time $T$ no column is unabsorbed.

For $|u|\le1/2$, $1+u+u^2\ge e^u$. Here $|\lambda Y|\le1/2$.
The denominator of~\eqref{eq:step} is at most $e^{\lambda^2 V}$ and we
rounded upward. Over all steps $\sum_t Y_t=k\langle h,\sigma\rangle$,
since each column ends at $k\sigma_j$. Applying the final score bound to
both signs gives
\[
 |\langle h,\sigma\rangle|
 \le \frac{2A\ln(4R)}{k}+\frac{TV}{2Ak}
 \le \frac{A}{8L}+O(L^3).
\]
Here the choice of $k$ bounds the first term, and $V\le2A$ and
$T/k=O(L^3)$ bound the second.

For the resource bound, $k=O(L^2)$ and $T=O(L^5)$. At a fixed time
the table entries are dyadic with a common denominator $2^{T-t}$.
Even one processor can build the whole table in $O(kT^2)$ elementary
steps by adding $O(T)$-bit numerators. There are $O(s)$ affine trials.
We give each trial a processor per sparse entry, row, and column.
The integer $JP_\pm$ is polynomially bounded by the score estimate, also
up to a constant factor for trial values. In a row update, multiplying
numerator and denominator by $4A^2$ reduces the computation of $JP'_\pm$
to word-size integer products and a rounded division.

In a score comparison, row terms have common denominator $J$ and
walk terms have a common power of two. We sum each group first, then
scale once. The resulting numerators have $O(T+L)$ bits. Additions and
comparisons take $O(T+L)$ elementary steps per worker, and multiplication
by an $O(L)$-bit integer takes $O((T+L)L)$.
There are $O(L)$ levels of reductions, including the minimum over trials.
Thus $O((T+L)L^2+L^3)$ steps per round suffice, and the total, including
precomputation, is $O(L^{12})$. Separate cells for the trials and a
minimum reduction implement every write on CREW.
\end{proof}

\section*{AI use statement.}
In developing this work, the authors used OpenAI Astra, Anthropic Opus~5.5,
and Google Argon to explore candidate constructions,
formalize proof sketches, and assist with writing and revision.
Our progress unfolded in three stages.
Starting from the prior randomized $\widetilde{O}(\Delta^2)$ depth bound,
we first obtained a deterministic $\widetilde{O}(\Delta^2)$-depth algorithm
by derandomizing parallel recoloring phases.
We next developed a polylogarithmic-depth single-star restoration routine that,
combined with vertex deletion, gave an $\widetilde{O}(n/\Delta)$ bound in the
high-degree regime and broke the $o(n)$ barrier at $\widetilde{O}(n^{2/3})$.
Finally, extending restoration from one star at a time to many stars in parallel reduced the degree dependence to deterministic $\widetilde{O}(\Delta)$.
The authors directed the process and evaluated the proposed ideas and arguments.

\bibliographystyle{plainnat}
\bibliography{ref}
\end{document}